\documentclass[a4paper,11pt,oneside]{article}

\usepackage{geometry} 
\usepackage{amsmath}
\usepackage{amssymb}
\usepackage{color}
\usepackage{graphicx}
\usepackage{subfig}
\usepackage{booktabs}
\usepackage{axodraw}
\usepackage{float}
\usepackage{afterpage}

\def\hbar{\hspace{0pt}\raisebox{1pt}{$-$} \hspace{-7pt} h}

\newcommand{\be}{\begin{equation}}
\newcommand{\ee}{\end{equation}}
\newcommand{\bea}{\begin{eqnarray}}
\newcommand{\eea}{\end{eqnarray}}
\newcommand{\nn}{\nonumber}

\newcommand{\TeV}{\,\mathrm{TeV}}
\newcommand{\GeV}{\,\mathrm{GeV}}

\newcommand{\keV}{\,\mathrm{keV}}

\newcommand{\nohyphens}%
      {\hyphenpenalty=10000\exhyphenpenalty=10000\relax}

\DeclareMathOperator{\re}{Re}
\DeclareMathOperator{\tr}{Tr}

\allowdisplaybreaks[1]

\newlength{\myem}
\newcounter{mysubequation}[equation]

\newcommand{\SISSA}{SISSA/ISAS and INFN, I--34151 Trieste, Italy}

\newcommand{\titletext}{\Large Minimal Yukawa-Gauge Mediation} 
\newcommand{\authortext}{\large Federica Bazzocchi\footnote{email: fbazzo@sissa.it}  \hspace{.1cm} and \hspace{.08cm} Maurizio Monaco\footnote{email: mmonaco@sissa.it}
\medskip\\\em\normalsize 
 \SISSA
\\[0.1\baselineskip] 
}
\newcommand{\abstracttext}{We consider a scenario in which Supersymmetry breaking is communicated to the MSSM fields through the interplay of yukawa and gauge interactions. The MSSM spectrum resembles that of split SUSY scenarios, but on top of that it develops some peculiar features like heavy higgsinos and an inverted hierarchy of sfermion masses. The predictions obtained are consistent with the most recent LHC SUSY and Higgs boson searches.}

\title{
\normalsize
\hspace*{\fill}
\vspace{3\baselineskip}\\\Large\bfseries\titletext\bigskip}
\author{\begin{minipage}[t]{0.8\textwidth}
\normalsize\centering\authortext
\end{minipage}}
\date{}

\begin{document}

 \maketitle
 \begin{abstract}\normalsize\noindent
 \abstracttext
 \end{abstract}\normalsize\vspace{\baselineskip}

 \clearpage


\tableofcontents


\section{Introduction}
In the last decades Supersymmetry (SUSY) has been considered as one of the most appealing extensions of the Standard Model (SM). For sure the presence of superpartners has provided one of the most elegant way to solve the so called  hierarchy problem associated to the SM Higgs boson mass \cite{susy}. The Large Hadron Collider (LHC) has been built  with the aim of investigating those energies at which new physics is expected. Historically SUSY was awaited at TeV scale in order to relieve the problem of fine tuning on EW scale. However the most recent results have been showing that SUSY, if it does exist, is not so close as previously expected. The community is now wondering which possibilities are still open to consider it as the correct extension of the SM.

In the last months LHC SUSY searches have pushed the lower bounds on colored sparticles up to the TeV scale. ATLAS \cite{ATLAS} and CMS \cite{CMS} most recent results have showed that both squarks and gluinos must be relevantly heavier than what expected in the pre-LHC era. Even if such analyses have been performed in specific frameworks, such as the CMSSM, and under specific hypothesis, doubtless at present we do not expect colored superpartners lighter than $1 \TeV$. What SUSY searches have not specifically excluded so far is the possible presence of low energy neutralinos or gravitino, for which the strictest bounds arise from direct and indirect  detection of Dark Matter (DM) \cite{DMall}.

The most famous LHC goal is the  Higgs boson discovery. The MSSM predicts 1 charged and  3 neutral, 2 CP even and 1 CP odd, scalars. The lightest neutral CP even scalar corresponds to the SM Higgs boson. The combined recent ATLAS and CMS results based on $2 \, \text{fb}^{-1}$ have shown that the windows from 146 GeV to 232 GeV, 256 GeV to 282 GeV and 296 GeV to 466 GeV are excluded at 95\% CL. A SM Higgs boson heavier than 466 GeV requires the presence of new physics -- typically new fermions -- to accommodate  electroweak precision tests (EWPT) and is not compatible with any SUSY framework.  In conclusion the solely window 115 - 146 GeV is still open for a SM-like Higgs and, if SUSY is the correct SM extension, the Higgs boson has to be found in this range.

The most recent analysis on LHC data have shown that the preferred hypothetical SUSY spectrum is quite peculiar and points in the direction of a scenario close to split SUSY \cite{ArkaniHamed:2004fb} or high scale SUSY \cite{Giudice:2011cg}. Indeed  these frameworks may accomodate heavy colored sparticles, relatively light neutralinos and a Higgs boson mass around 135 GeV as it has been already noticed in some recent papers \cite{Alves:2011ug}. What at the moment seems quite clear is that if SUSY exists the low scale masses are obtained by a quite large tuning of the parameters. This means that the fine tuning principle has to be reassessed if taken as a guideline in model building.

Here we present a  new  SUSY breaking framework  in which the  spectrum so far sketched is naturally achieved. In our  scenario SUSY breaking is communicated via the combined effect of yukawa and gauge mediation.  In particular we revisit in a minimal version the old idea of yukawa-gauge mediation \cite{Chacko:2001km}. The paper is organized as follows:  in section \ref{model} we describe the mechanism  of  SUSY breaking mediation firstly with a toy model and then in the MSSM  context. In section \ref{pheno} we show how the soft terms arise and we focus on the phenomenological predictions of the model. We discuss in detail the  sparticle spectrum, the Higgs boson mass and EWPT. We also  show that  the model is safe with respect to flavor changing neutral current (FCNC) processes and we give some cosmological considerations. Section \ref{outlook} is devoted to the outlook, while section \ref{conc} to our conclusions.

\section{The model}
\label{model}

As anticipated the model we are going to propose is a simpler version of yukawa-gauge mediation  scenarios already present in literature \cite{Chacko:2001km}.  As we will see it is simpler because we do not ask for  a GUT completion and we use the MSSM Higgs doublets to act as SUSY breaking messengers. In this way we avoid the use of large representations and the number of new superfields added to the MSSM ones is really basical.

\subsection{General implant: a toy model}
\label{toy}

In this section we use a toy model to introduce our SUSY breaking mechanism. Let us consider a SUSY gauge theory based on the simple group $\mathcal{G}$. Matter superfields  are charged under $\mathcal{G}$ and  denoted by $\hat{Q}_i$. Here we  do not address the origin of SUSY breaking, assuming it happens because of an unknown mechanism in a secluded sector. For our purposes we just consider that the net effect of such a breaking can be parametrized by a gauge singlet chiral superfield $\hat{X} = X + \theta \psi_X + \theta^2 F_X$ developing vev in its scalar $\langle X\rangle  = M / k$ and auxiliary $\langle {F_X}\rangle = F / k$ components. The coupling constant $k$, whose meaning will soon be apparent, is introduced in the vev definition for later convenience. Such a chiral superfield cannot mediate SUSY breaking, thus we have to couple it to a  charged superfield sector that effectively communicates SUSY breaking to matter. At this level the scenario is similar to minimal gauge mediation \cite{Giudice:1998bp}: the field that develops a SUSY breaking vev does not couple directly to MSSM fields. However, contrary to minimal gauge mediation, $\hat{X}$ couples only to an additional gauge singlet  $\hat{\Phi}$, and the latter interacts with charged superfields, thus our mechanism works through two messenger sectors and therefore in two different steps. $\hat{\Phi}$ is identified as first messenger: it  couples to charged superfields,  $\hat{H}_i$ (and their partners with opposite charge $\hat{\overline{H}}_i$), that effectively perform the SUSY breaking mediation. The superpotential that implements the two step mechanism just sketched has the form
\begin{equation}
W_{messengers} = k \hat{X} \hat{\Phi} \hat{\Phi} + \lambda_{ij} \hat{\Phi} \hat{H}_i \hat{\overline{H}}_j.
\end{equation}
The toy model superpotential contains  also  the mass term for the second messengers
\begin{equation}
W_{mass} = \mu_{ij} \hat{H}_i \hat{\overline{H}}_j \, .
\end{equation} 
While the first messenger, $\hat{\Phi}$, is typically thought quite heavy and it decouples from the low energy spectra, the $\hat{H}_i$ superfields could be decoupled or not according to the structure of $\mu_{ij}$. In practice we will see that  in our realization a subset of the $\hat{H}_i$ fields  becomes part of the low energy spectrum. 

\begin{figure}[h]
\begin{center}
\begin{picture}(108,63)(-54,-27)
\SetWidth{0.9}
\Photon(-55,0)(-22,0){3}{3}
\Line(-55,0)(-22,0)
\CArc(0,0)(22,0,180)
\DashCArc(0,0)(22,180,0){5}
\CArc(0,0)(4,0,360)
\Line(-2.8,-2.8)(2.8,2.8)
\Line(2.8,-2.8)(-2.8,2.8)
\DashLine(0,-22)(0,-4){4}
\DashLine(0,4)(0,22){4}
\Photon(55,0)(22,0){3}{3}
\Line(55,0)(22,0)
\Text(-45,9)[c]{$\lambda$}
\Text(45,9)[c]{$\lambda$}
\Text(-23,23)[c]{$\psi_H$}
\Text(23,23)[c]{$\psi_H$}
\Text(-23,-23)[c]{$H$}
\Text(23,-23)[c]{$H$}
\Text(-5,12)[c]{$\phi$}
\Text(-5,-12)[c]{$\phi$}
\Text(11,0)[c]{$B_\phi$}
\Text(0,-34.5)[c]{(a)}
\end{picture}
\hspace{1.4cm}
\begin{picture}(108,63)(-54,-27)
\SetWidth{0.9}
\DashLine(-55,0)(-22,0){4}
\DashCArc(0,0)(22,-75,255){5}
\CArc(0,-22)(4,0,360)
\Line(-2.8,-24.8)(2.8,-19.2)
\Line(2.8,-24.8)(-2.8,-19.2)
\DashLine(22,0)(45.1,23.1){4}
\DashLine(22,0)(45.1,-23.1){4}
\Text(-45,9)[c]{$H$}
\Text(45,15)[c]{$Q$}
\Text(45,-15)[c]{$Q$}
\Text(0,28)[c]{$H$}
\Text(-13,-8)[c]{$\phi$}
\Text(13,-8)[c]{$\phi$}
\Text(0,-12)[c]{$B_\phi$}
\Text(0,-34.5)[c]{(b)}
\end{picture}
\end{center}
\vspace{0.1cm}
\begin{center}
\begin{picture}(108,63)(-54,-27)
\SetWidth{0.9}
\DashLine(-55,0)(-38.5,0){4}
\Line(-38.5,0)(-22,0)
\CArc(0,0)(38.5,0,180)
\CArc(0,0)(22,0,180)
\DashCArc(0,0)(22,-180,-130){5}
\DashCArc(0,0)(22,-110,-70){5}
\DashCArc(0,0)(22,-50,0){5}
\CArc(-11,-19.5)(4,0,360)
\CArc(11,-19.5)(4,0,360)
\Line(-7.3,-22.4)(-13.8,-16.6)
\Line(-7.3,-16.6)(-13.8,-22.4)
\Line(7.3,-22.4)(13.8,-16.6)
\Line(7.3,-16.6)(13.8,-22.4)
\Line(22,0)(38.5,0)
\DashLine(38.5,0)(55,0){4}
\Text(-48,9)[c]{$Q$}
\Text(48,9)[c]{$Q$}
\Text(-29,6)[c]{$\psi_H$}
\Text(29,6)[c]{$\psi_H$}
\Text(0,32)[c]{$\psi_Q$}
\Text(0,15)[c]{$\psi_\phi$}
\Text(-16,-5)[c]{$\phi$}
\Text(0,-15)[c]{$\phi$}
\Text(16,-5)[c]{$\phi$}
\Text(14,-30)[c]{$B_\phi$}
\Text(-14,-30)[c]{$B_\phi$}
\Text(0,-38.5)[c]{(c)}
\end{picture}
\hspace{1.4cm}
\vspace{0.3cm}
\caption{Loop graphs giving rise to the different soft terms: (a) gaugino masses, (b) trilinear terms, (c) sfermion masses.
\label{fig:toydiagram}}
\end{center}
\end{figure}
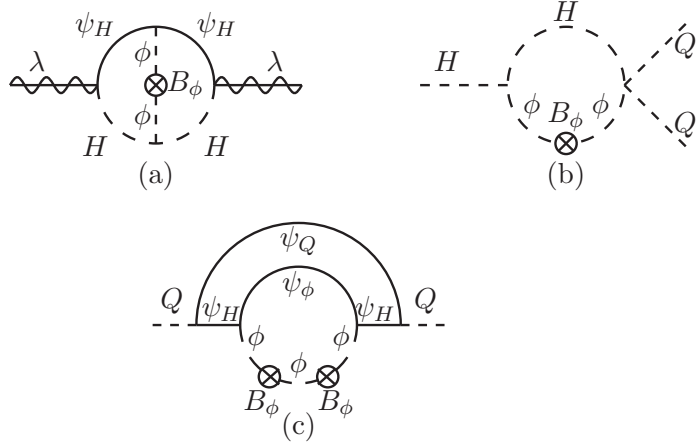

The $\hat{H}_i$ superfields are  the effective mediators of SUSY breaking to  the MSSM superfields.  Gaugino masses are given by two loop graphs in which a gaugino goes through gauge interactions to $\hat{H}_i$ and then the latter couples to the $\hat{\Phi}$ superfield loop, as shown in figure \ref{fig:toydiagram}. The effects of SUSY breaking to the matter sector are mediated by means of superpotential trilinear interactions
 \begin{equation}
W_{matter} = h_{ijk} \hat{Q}_i \hat{Q}_j \hat{H}_k + \overline{h}_{lmn} \hat{Q}_l \hat{Q}_m \hat{\overline{H}}_n\,,
\end{equation} 
where the indices $(ijk)$ and $(lmn)$ are contracted to give gauge invariants. In this scenario the SUSY breaking trilinears are induced at one loop level while sfermion masses arise at two loops as shown in figure \ref{fig:toydiagram}. 

\subsection{General implant: explicit construction of the model}

Now we show how the scenario proposed can be implemented to describe particle physics phenomenology. The ingredients are essentially those  anticipated in the  toy model. 

The effect of the secluded sector where SUSY breaking effectively takes place are just parametrized by the presence of the gauge singlet chiral superfield $\hat{X} = X + \theta \psi_X + \theta^2 F_X$ that develops a vev both in its scalar  and auxiliary  components, $M/k$ and $F/k$ respectively. $\hat{X}$ couples to a gauge chiral singlet $\hat{\Phi}$ through a superpotential term $k \hat{X} \hat{\Phi} \hat{\Phi}$.  The gauge group $\mathcal{G}$ is the SM gauge group and we identify the second messenger sector $(\hat{H}_i,\hat{\overline{H}}_i)$  with the MSSM Higgs superfields. In order to prevent too heavily suppressed gluino masses we can add an extra heavy colored triplet with the quantum numbers of the down quark superfield, that does not couple dangerously to MSSM fields by means of a $Z_2$ discrete symmetry. Thus the second messengers of the framework happen to be $(\hat{H}_i,\hat{\overline{H}}_i) = (\hat{H}_u\oplus \hat{T}, \hat{H}_d \oplus \hat{\overline{T}})$.  In this way  all the three gauginos receive mass at the same loop level.  Clearly the matter fields $\hat{Q}_i$ are the MSSM matter superfields. 

The extra triplet must be heavy, but not necessarily of order the GUT scale. Dangerous operators mediating the proton decay are actually forbidden by the unbroken $Z_2$ symmetry. In order to correctly mediate SUSY breaking the triplet have not to be decoupled when $\hat{X}$ gets vev. We can set a lower bound on $M_T$ asking for gauge coupling unification to be achievable in the scenario. The model  field content is  contained in table \ref{tab:content}.
\begin{table}[h!]
\caption{Superfield content of the theory.}
\begin{center} 
\begin{tabular}{|c|c|c|c|c|c|} 
\hline \hline 
SF & Spin 0 & Spin \(\frac{1}{2}\) & Generations & \((U(1)\otimes\, \text{SU}(2)\otimes\, \text{SU}(3))\) \\ 
\hline 
\(\hat{q}\) & \(\tilde{q}\) & \(q\) & 3 & \((\frac{1}{6},{\bf 2},{\bf 3}) \) \\ 
\(\hat{l}\) & \(\tilde{l}\) & \(l\) & 3 & \((-\frac{1}{2},{\bf 2},{\bf 1}) \) \\ 
\(\hat{H}_d\) & \(H_d\) & \(\tilde{H}_d\) & 1 & \((-\frac{1}{2},{\bf 2},{\bf 1}) \) \\ 
\(\hat{H}_u\) & \(H_u\) & \(\tilde{H}_u\) & 1 & \((\frac{1}{2},{\bf 2},{\bf 1}) \) \\ 
\(\hat{d}\) & \(\tilde{d}^c\) & \(d^c\) & 3 & \((\frac{1}{3},{\bf 1},{\bf \overline{3}}) \) \\ 
\(\hat{u}\) & \(\tilde{u}^c\) & \(u^c\) & 3 & \((-\frac{2}{3},{\bf 1},{\bf \overline{3}}) \) \\ 
\(\hat{e}\) & \(\tilde{e}^c\) & \(e^c\) & 3 & \((1,{\bf 1},{\bf 1}) \) \\ 
\(\hat{\Phi}\) & \(\phi\) & \(\psi_{\Phi}\) & 1 & \((0,{\bf 1},{\bf 1}) \) \\ 
\(\hat{T}\) & \(t\) & \(\psi_t\) & 1 & \((\frac{1}{3},{\bf 1},{\bf \overline{3}}) \) \\ 
\(\hat{\overline{T}}\) & \(\overline{t}\) & \(\psi_{\overline{t}}\) & 1 & \((-\frac{1}{3},{\bf 1},{\bf 3}) \) \\
\hline \hline
\end{tabular} 
\end{center} 
\label{tab:content}
\end{table}

At the GUT scale the superpotential is given by two contributions
\begin{equation}
W= W_{MSSM}+ W_\Phi\,,
\end{equation}
with 
\begin{eqnarray} 
&& W_{MSSM} =    Y_u\,\hat{q}\,\hat{H}_u\,\hat{u}\,- Y_d \,\hat{q}\,\hat{H}_d\,\hat{d}\,- Y_e \,\hat{l}\,\hat{H}_d\,\hat{e}\,+\mu\,\hat{H}_u\,\hat{H}_d\,,\nn\\
&& W_{\Phi} =  
h_0\,\hat{H}_u\,\hat{H}_d\,\hat{\Phi} +\frac{1}{3} \eta \,\hat{\Phi}\,\hat{\Phi}\,\hat{\Phi}\,+k \hat{X} \,\hat{\Phi}\,\hat{\Phi}\, +M_T\,\hat{T}\,\hat{\overline{T}}\,+h_t\,\hat{T}\,\hat{\overline{T}}\,\hat{\Phi}.
\end{eqnarray} 
As anticipated the extra $Z_2$ discrete symmetry prevents the triplets to couple with matter fields.  When $\hat{X}$ develops vev and breaks SUSY, $\hat{\Phi}$ communicates such a breaking to the MSSM fields giving rise to the standard MSSM soft potential 
\begin{eqnarray}
\label{vsoft}
V_{soft}&=&m^2_{H_u} |H_u|^2+m^2_{H_d} |H_d|^2+m^2_{Q_i} |\tilde{Q}|^2+ m_{u}^{2} |\tilde{u}^c|^2\nn\\
&+& m_{d}^{2} |\tilde{d}^c|^2 + m_{L}^{2} |\tilde{L}|^2 + m_{e}^{2} |\tilde{e}^c|^2 \nonumber \\ 
 &+&  H_u \tilde{u}^c \tilde{Q} A_{u} +H_d \tilde{d}^c \tilde{Q} A_{d} + H_d \tilde{e}^c \tilde{L} A_{e}\,,
\end{eqnarray}
that is summed to the SUSY invariant scalar potential 
\begin{equation}
V_{SUSY}=\left| \Biggl(\frac{\partial W_{MSSM}}{\partial \hat{\Omega}_j} \bigg\vert_{\hat{\Omega}_j=\tilde{\Omega}_j}\Biggr) \right|^2+ \sum_a \left| \tilde{\Omega}^{\dagger}_j T_a \tilde{\Omega}_j\right|^2 \,,
\end{equation}
where $\hat{\Omega}_j$ are the various superfields of the theory. Below the scale $M$ ($M_T$) $\hat{\Phi}$ ($\hat{T}, \hat{\overline{T}}$) decouple and we are left  with the standard MSSM. For this reason we have not included terms involving these fields in equation (\ref{vsoft}). We neglect threshold effects arising from the decoupling of these heavy fields. The soft terms structure is strongly correlated to all the superpotential parameters, because of the yukawa-gauge mediation mechanism. This in turn gives rise to a novel spectrum, that is the subject of next sections.

\section{Phenomenological predictions}
\label{pheno}
This section is devoted to the phenomenological predictions of our model. The mass spectrum is peculiar to this singlet yukawa-gauge mediation realization and quite different from that obtained in minimal gauge mediation   frameworks. 

\subsection{Spectrum}
\label{spectrum}
We have  already anticipated  that the spectrum of  the theory is quite uncommon: indeed the third family sfermions are in general heavier than those of the first two families, a feature owed to the role played by the yukawas in the SUSY breaking mediation mechanism. A similar hierarchy has been recently considered in \cite{Endo:2010fk}. The structure of the low energy spectrum is determined by the RG evolution of the boundary contributions generated when integrating out the first messenger, $\hat{\Phi}$, at its scale $M$. In the following sections we will show that in order to be phenomenologically acceptable our model requires $M \lesssim 10^{14}$. Thus for simplicity in the following we assume that $M \sim M_T < M_{GUT}$ and leave the possibility  of a  low energy SUSY breaking realization to further studies. In particular it could be interesting to connect the superfield $\hat{\Phi}$ to the generation of neutrino masses \cite{inprogress}. In addition we will assume that the democratic contribution to sfermion mass matrices arising because of the gravitino is negligible. We will comment on this assumption in section \ref{cosmology}. Moreover we consider anomaly mediation contributions \cite{anomaly} to be subleading.

To deeper analyze the implant of the theory and the differences with respect to standard scenarios  we should remember that the MSSM fields couple in a quite peculiar way to the SUSY breaking vev. In particular at one loop level no soft mass terms are generated. The first contributions  appear at two loops, where both gauginos and sfermions get a mass term. The structure of the two terms (see section \ref{soft}) is 
\begin{equation}
\label{2loopcoeff}
M_{gaugino} = Lp^2 A^{i,j} g_i^2 h_j^2 B_{\phi}\,,\qquad \qquad m^2_{sfermion} = Lp^2 B^{r}\mathcal{Y}_{r} h_0^2  B_{\phi}^2 \,,
\end{equation} 
where $B_\phi= F/M \ll M$, $Lp = (4 \pi)^{-2}$ is a loop factor, $\mathcal{Y}_{u(d)}= {Y}^\dag_{u(d)} Y_{u(d)},  Y_{u(d)} {Y}^\dag_{u(d)}$ for right and left up (down) quark respectively and $h_j = h_0 ,\, h_t$. With respect to minimal gauge mediation there is an effective extra loop factor in gaugino masses, keeping them smaller than third family sfermions. Indeed sfermion masses are proportional to the yukawa couplings and thus they result heavily suppressed in the case of the first two families. Anyway the contribution of minimal gauge mediation coming from three loop graphs become competing or even more important of the yukawa mediated two loop one in this case. Such a contribution yields terms of the form
\begin{equation}
\label{3loopcoeff}
m^2_{sfermion} = Lp^3 \Bigl( C^{i,j,k}_1 r^2_i r^2_j r^2_k + C^{i,j,k}_2 r^2_i r^2_j \mathcal{Y}_k + C^{i,j,k}_3 r^2_i \mathcal{Y}_j \mathcal{Y}_k \Bigr) B_{\phi}^2\,,
\end{equation}
where $r_i = g_{j=1,2,3}, \, h_0, \, h_t, \, \eta$. The three loop contributions happen to be competing with the two loop one only in the case of third family down and lepton sfermions, that are characterized by small yukawa couplings compared to the top one. On the contrary  they are dominant in the case of the first two families for all the flavours, since in that case  the yukawa couplings give rise to negligible terms. Consequently the first two families are essentially degenerate in mass, a feature  preserved even after the evolution to low energies. For what concerns the first two families the hierarchy with respect to  gaugino masses is milded because of the extra loop factor, but still present. 

Below the scale $M$ we are left with the particle content of the MSSM, thus the evolution of soft terms can be simply obtained using the $\beta$ functions reported in \cite{RGE}. 

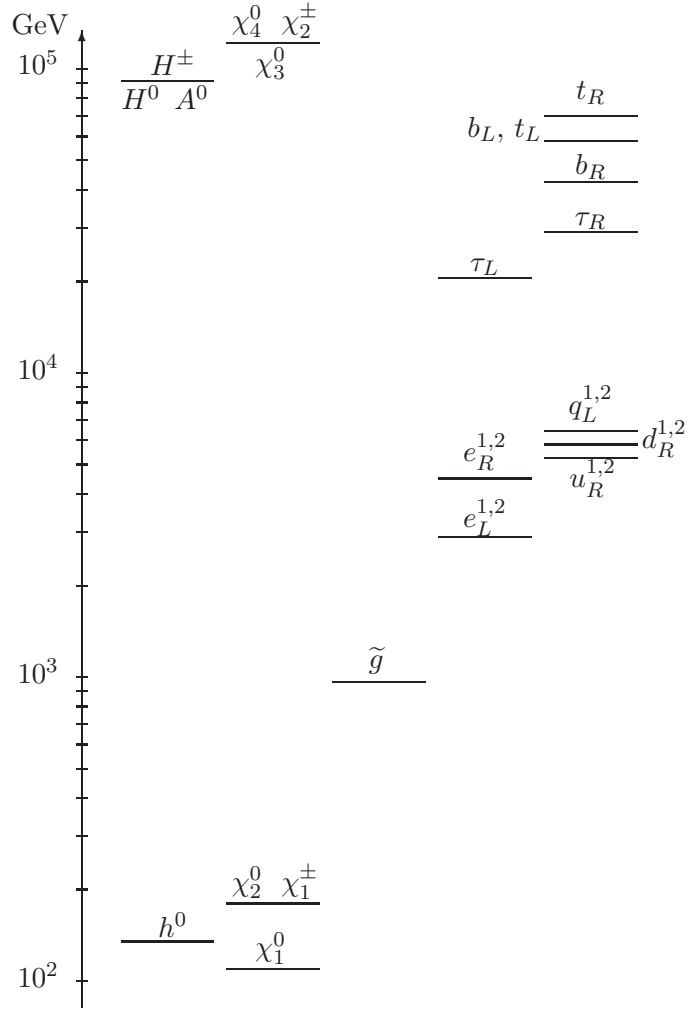
\begin{figure}
\begin{center}
\begin{picture}(250,375)(0,0)
\put(25,-10){\vector(0,1){370}}
\put(23,0){\line(1,0){4}}
\put(6,0){\makebox(5,5){$10^2$}}
\put(23,34.66){\line(1,0){4}}
\put(23,54.93){\line(1,0){4}}
\put(23,69.31){\line(1,0){4}}
\put(23,80.47){\line(1,0){4}}
\put(23,89.59){\line(1,0){4}}
\put(23,97.29){\line(1,0){4}}
\put(23,103.97){\line(1,0){4}}
\put(23,109.86){\line(1,0){4}}
\put(23,115.13){\line(1,0){4}}
\put(6,115.13){\makebox(5,5){$10^3$}}
\put(23,149.78){\line(1,0){4}}
\put(23,170.06){\line(1,0){4}}
\put(23,184.44){\line(1,0){4}}
\put(23,195.601){\line(1,0){4}}
\put(23,204.717){\line(1,0){4}}
\put(23,212.425){\line(1,0){4}}
\put(23,219.101){\line(1,0){4}}
\put(23,224.99){\line(1,0){4}}
\put(23,230.26){\line(1,0){4}}
\put(6,230.26){\makebox(5,5){$10^4$}}
\put(23,264.92){\line(1,0){4}}
\put(23,285.19){\line(1,0){4}}
\put(23,299.57){\line(1,0){4}}
\put(23,310.73){\line(1,0){4}}
\put(23,319.85){\line(1,0){4}}
\put(23,327.55){\line(1,0){4}}
\put(23,334.23){\line(1,0){4}}
\put(23,340.12){\line(1,0){4}}
\put(23,345.39){\line(1,0){4}}
\put(6,345.39){\makebox(5,5){$10^5$}}
\put(6,360){\makebox(5,5){$\GeV$}}
\put(40,15.01){\line(1,0){35}}
\put(57,19){\makebox(5,5){$h^0$}}
\put(40,340.8){\line(1,0){35}}
\put(45,332.15){\makebox(5,5){$H^0$}}
\put(40,340.8){\line(1,0){35}}
\put(64,332.15){\makebox(5,5){$A^0$}}
\put(40,340.8){\line(1,0){35}}
\put(57,345.15){\makebox(5,5){$H^\pm$}}
\put(80,4.76){\line(1,0){35}}
\put(94,10){\makebox(5,5){$\chi^0_1$}}
\put(80,29.38){\line(1,0){35}}
\put(85,35){\makebox(5,5){$\chi^0_2$}}
\put(80,355.14){\line(1,0){35}}
\put(94,345){\makebox(5,5){$\chi^0_3$}}
\put(80,355.14){\line(1,0){35}}
\put(85,361){\makebox(5,5){$\chi^0_4$}}
\put(80,29.39){\line(1,0){35}}
\put(105,35){\makebox(5,5){$\chi^\pm_1$}}
\put(80,355.14){\line(1,0){35}}
\put(105,361){\makebox(5,5){$\chi^\pm_2$}}
\put(120,113.28){\line(1,0){35}}
\put(134,119){\makebox(5,5){$\widetilde{g}$}}
\put(160,266.08){\line(1,0){35}}
\put(175,268){\makebox(5,5){$\tau_L$}}
\put(200,283.49){\line(1,0){35}}
\put(215,285.5){\makebox(5,5){$\tau_R$}}
\put(200,302.63){\line(1,0){35}}
\put(215,306){\makebox(5,5){$b_R$}}
\put(200,318.14){\line(1,0){35}}
\put(180,320){\makebox(10,5){$b_L$, $t_L$}}
\put(160,168){\line(1,0){35}}
\put(175,173){\makebox(5,5){$e_L^{1,2}$}}
\put(160,190.33){\line(1,0){35}}
\put(175,198){\makebox(5,5){$e_R^{1,2}$}}
\put(200,327.62){\line(1,0){35}}
\put(215,334){\makebox(5,5){$t_R$}}
\put(200,198.16){\line(1,0){35}}
\put(215,188){\makebox(5,5){$\,u_R^{1,2}$}}
\put(200,203.24){\line(1,0){35}}
\put(242,203){\makebox(5,5){$\,d_R^{1,2}$}}
\put(200,208.33){\line(1,0){35}}
\put(215,216){\makebox(5,5){$q_L^{1,2}$}}
\end{picture}
\end{center}
\caption{Typical spectrum arising in the framework. The parameters to obtain it are $h_0 = 1.04$, $h_t = 1.30$, $B_\phi = 9.7 \times 10^6 \GeV$, $\eta = 0.1$.}
\label{figure:spectra}
\end{figure}

Since our theory predicts the presence of a split spectrum in which sfermions and Higgsinos are much heavier than gauginos, we improved the calculation of gaugino masses by integrating out  sfermions and higgsinos  at their mass scales and then determining new evolution equations. Such a procedure is explained in details in next subsection to show that gauge coupling unification predictions are not affected by this spectrum. In figure \ref{figure:spectra} we report an example of low energy spectrum obtained within our framework.

\subsection{Gauge coupling unification}
\label{section:gaugeunification}
In this section we briefly discuss how gauge coupling unification is realized in  our scenario.  Notice that in our model unification is not mandatory. 

We take into account only the one loop RGEs\footnote{Two loop RG evolution, as well as one  loop threshold corrections, give non negligible contribution, but they even better unification, as discussed in appendix \ref{appendix:gaugeunification}.} with a series of intermediate scales dictated  by the  spectrum shown in figure \ref{figure:spectra}. Here we just report the main results while all the detailed calculations can be found in appendix \ref{appendix:gaugeunification}.

\begin{figure}
\centering
\subfloat[Evolution of gauge couplings]{\label{figure:gauge_evolution}\includegraphics[width=0.45\textwidth]{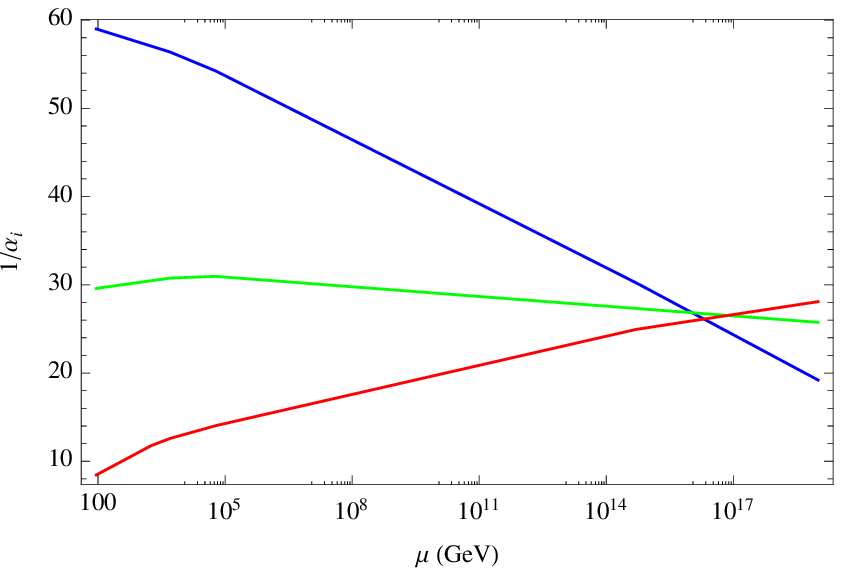}}
\subfloat[Unification scale under a magnifying glass]{\label{figure:unification}\includegraphics[width=0.481\textwidth]{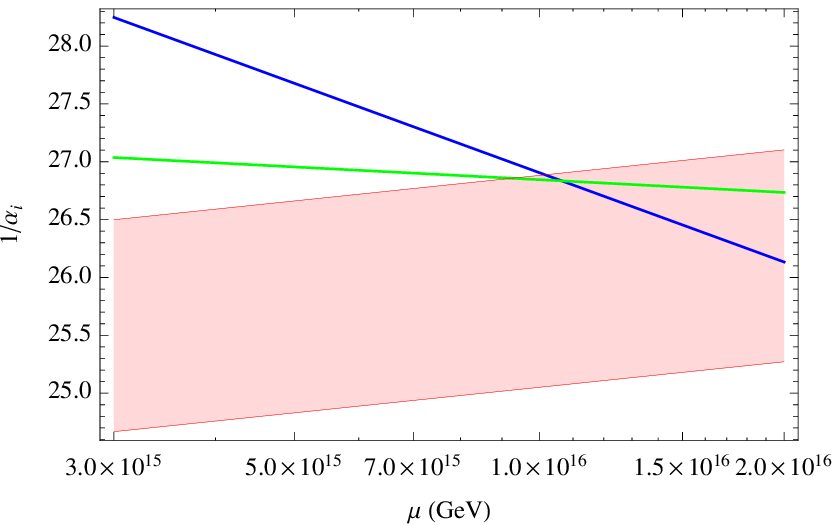}}
\caption{Gauge coupling evolution.} 
\end{figure}

Neglecting the electroweak scale, characterized  by the SM degrees of freedom,  in  the spectra of figure \ref{figure:spectra} we identify four SUSY scales:  the lower one, $M^{(4)}_{SUSY}\sim $ hundreds of GeV,  is that of $SU(2)_L \times U(1)_Y$ gauginos. Then follow the gluino scale,  $M^{(3)}_{SUSY}$,  that LHC constraints fix around the TeV, and the light sfermions one $M^{(2)}_{SUSY}$. The first scale, $M^{(1)}_{SUSY}$, corresponds to the   the heavy third family sfermions, higgsinos and heavy scalars. Finally we have to consider the scale of the heavy triplets $M_T$. The evolution is then computed taking as inputs the low energy gauge coupling values and  the result is shown in figure \ref{figure:gauge_evolution}. In order to test gauge coupling unification we show in figure \ref{figure:unification} a plot at energies around unification scale under the magnifying glass. The red strip represents the region for the strong coupling within three sigmas of the experimental value; thus we can easily see that it is compatible with unification if we consider that two loop contributions better the picture, as explained in appendix \ref{appendix:gaugeunification}. 

\subsection{EWSB and a light Higgs boson}
\label{higgs}

It is well known \cite{Giudice:2011cg} that even in the presence of high scale SUSY the lightest neutral scalar remains relatively light and still compatible with the narrow light-mass window left open by the recent exclusion plots. While waiting for the new analysis to see if this window  will disappear or not,  discussing scenarios predicting a Higgs boson around 130 GeV is still reasonable.

We said  that the spectrum is essentially divided in two different sets. While the gauginos are expected to lie at lower energy ($M^{(4,3)}_{SUSY}\gtrsim 100 \GeV \div \TeV$), all the other sparticles (namely the sfermions and the Higgsinos) are pushed up to multi-TeV energies, $M^{(2,1)}_{SUSY}$. Such a spectrum might be easily confused with a split SUSY one, but it actually differs from that because in our picture the Higgsinos are  heavy  and the lightest neutralino is essentially a bino or wino. 

The light Higgs boson mass is obtained by the standard procedure used in SUSY scenarios, by decoupling heavy particles in turn and computing their threshold effects as the energy decreases. Discussing gauge coupling unification we identified four scales at which sequentially heavy particles decouple. At the highest scale below GUT scale, $M_{SUSY}^{(1)}$, we decouple third family sfermions,  Higgsinos and  heavy Higgs scalars. Indeed only one linear combination of $H_u$ and $H_d$ 
\begin{equation}
\label{light}
h = H_d \cos \alpha + i \sigma_2 H_u^* \sin \alpha,
\end{equation}
remains part of the theory, corresponding to  the light Higgs boson. The orthogonal combination 
\begin{equation}
\label{heavy}
H = - H_d \sin \alpha + i \sigma_2 H_u^* \cos \alpha,
\end{equation}
corresponds to  an $SU(2)_L$ heavy doublet, whose mass is roughly  fixed by the soft terms. The two masses are given at $M^{(1)}_{SUSY}$ by
\begin{eqnarray}
\label{masses}
m^2_{h,H}(M^{(1)}_{SUSY})&=& \frac{1}{2}\left[ m^2_{H_u}+m^2_{H_d}+ 2|\mu|^2\pm  (m^2_{H_d}-m^2_{H_u})/\cos 2 \alpha\right ]\,.
\end{eqnarray}
It is clear that  $m^2_h$ is the quadratic term of the Higgs scalar potential that has to run to negative values at the EW scale to induce EWSB. On the contrary $m^2_H$ has to be greater than zero being the mass of the heavy doublet. $\alpha$ in equations (\ref{light}), (\ref{heavy}) and (\ref{masses}) is meant to be the mixing angle between the light and heavy states before EWSB, defined by
\begin{equation}
\tan 2 \alpha= -\frac{2 B_\mu}{(m^2_{H_u}-m^2_{H_d})}\,.
\end{equation}
The light state coincides roughly with the SM Higgs boson and it is responsible of EWSB. Its mixing with $H$ induces a negligible vev $v_H$, thus the mixing angle $\alpha$ and the ratio $\beta = v_u/v_d$ do essentially coincide. In our framework $B_\mu$ is tightly connected to $\mu$  and to the soft $m^2_{H_{u,d}}$, in particular the SUSY breaking contributions at $M$ are
 \begin{eqnarray}
B_\mu &\sim& Lp \, h_0^2 \, \mu \, B_\phi\,, \nn\\
m^2_{H_{u,d}} &\sim&  Lp^2 \, h_0^2 \, \Bigl(\frac{18}{5} g^2 -h_0^2-h_t^2 -2 \eta^2 - C_Y^{u,d}\Bigr)B_\phi^2\,,
\end{eqnarray}
where $Lp=(4 \pi)^{-2}$, $C_Y^u \sim y_\tau+3 y_b^2,C_Y^d \sim 3 y_t^2$ and we have assumed $g_1\sim g_2$ being close to the GUT scale. From
the large cancellation needed to get  $m_h^2$ we have  $ |\mu|\sim B_\phi/(16 \pi^2)$, thus
 \begin{equation}
 \tan 2 \beta  \sim 1/ C_Y^d\sim \mathcal{O}(1)\,,
 \end{equation} 
 since $C_Y^d \gg C_Y^u$.
Consequently the natural range is $\tan\beta < 1$, but this would destroy yukawa coupling  perturbativity. Thus to allow larger values for $\tan \beta$ we need a mechanism to increase $B_\mu$. We assume this is possible without modifying the main features of our scenario\footnote{This may be realized by introducing  higher order operators to $W_{MSSM}$ or  adding a scalar singlet that develops a vev as suggested in \cite{Chacko:2001km}.} and we do not discuss further details. 
 
The effective theory below $M_{SUSY}^{(1,2)}$  contains the  SM matter content and the three gauginos. The doublet h is nothing more  but the usual SM Higgs field 
whose scalar potential, $V(h)$, is characterized by the quartic coupling $\lambda$. As usual $\lambda$ is given by the SUSY  tree level contributions
\begin{equation}
\lambda_{SUSY} = \frac{1}{4} \Bigl(g_2^2(M_{SUSY}^{(1)}) + \frac{3}{5} g_1^2(M_{SUSY}^{(1)})\Bigr) \cos^2 2 \beta(M_{SUSY}^{(1)}),
\end{equation}
and the one-loop threshold contribution obtained integrating out  heavy sfermions via box and triangle one-loop diagrams  \cite{Okada:1990gg}. The dominant contribution arises from diagrams involving the stop, even if in our case the stop is the heaviest sfermion, and it is given by
\begin{equation}
\label{deltal}
\delta \lambda = \frac{3 Y_t^4(M_{SUSY}^{(1)})}{16 \pi^2} \left(2 \frac{X_t^2}{{M_{SUSY}^{(1)}}^2} - \frac{X_t^4}{6 {M_{SUSY}^{(1)}}^4} \right)\,,
\end{equation}
where $X_t = A_t(M_{SUSY}^{(1)}) + \mu(M_{SUSY}^{(1)}) \cot \beta(M_{SUSY}^{(1)})$.  

$\lambda$ is then evolved through the RGEs  given in appendix \ref{appendix:higgs} up to the EW scale, where the one loop effective Coleman Weinberg potential \cite{ColemanWeinberg} is computed

\begin{equation*}
V(h) = m^2_h |h|^2 + \frac{1}{2} \lambda |h|^4 + \frac{1}{16 \pi^2} \sum_{k=1}^5 a_k \mathcal{M}^4_k \left( \ln \frac{\mathcal{M}^2_k}{\mu^2} + b_k \right)
\end{equation*}
where
\begin{eqnarray*}
\mathcal{M}_1^2 = m_h^2 + \frac{\lambda}{2} h^2 \,, \qquad \mathcal{M}_2^2 = m_h^2 + \frac{\lambda}{6} h^2 \,, \qquad \mathcal{M}_3^2 = \frac{1}{4} g_2^2 h^2 \,, \\
\mathcal{M}_4^2 = \frac{1}{4} \Bigl(g_2^2 + \frac{3}{5} g_1^2\Bigr) h^2 \,, \qquad \mathcal{M}_5^2 = m_h^2 + \frac{h_t^2}{2} h^2 \,,
\end{eqnarray*}
and
\begin{eqnarray*}
a_1 = \frac{1}{4} \,, \qquad a_2 = \frac{3}{4} \,, \qquad && a_3 = -3 \,, \qquad a_4 = \frac{3}{2} \,, \qquad a_5 = \frac{3}{4} \,, \\
b_1 = b_2 = b_3 = - \frac{3}{2} \,, &&\qquad b_4 = b_5 = - \frac{5}{6} \,.
\end{eqnarray*}

The procedure used to find the minimum of the Higgs potential is the one depicted in \cite{Casas_Quiros}. The running mass of the Higgs boson is defined as the second derivative of the potential evaluated at the minimum, namely
\begin{equation}
\label{rmass}
\hat{m}^2_h = \frac{\partial^2 V(h)}{\partial^2 h}\Bigg\vert_{\langle{h}\rangle=v_W} \,,
\end{equation}
where $v_W$ is the EW scale.
Finally the  physical Higgs boson mass is obtained by computing  the pole mass. The relation between the Higgs running mass and the pole one can be evaluated as follows. We can write 
\begin{equation}
M^2_h = \hat{m}^2_h + \Delta \Pi \,,
\end{equation}
where $M^2_h$ is the pole propagator mass, $\hat{m}^2_h$ is the running Higgs mass defined in (\ref{rmass})  and $\Delta \Pi$ is the difference of the renormalized self energy calculated at the pole mass and at zero momentum: $\Delta \Pi \equiv \Pi(p^2 = M^2_h) - \Pi(p^2 = 0)$. $\Delta \Pi$ receives contribution from many sources,  being the top contribution the most relevant.

In the parameter point corresponding to Figure \ref{figure:spectra} the pole mass of the light Higgs boson is 
\begin{equation}
M^2_h = 133 \GeV \,,
\end{equation}
that is inside the light mass window left open by the latest data. This value is in agreement with  the results obtained by \cite{Giudice:2011cg}. Clearly a full analysis of the parameter space could be interesting to see  the spread of our prediction.

\subsection{EWPT}

As in any theory providing a SM extension we have to check the consistence  of our model  through  the oblique corrections classified \cite{TSU1} by means of  the three parameters  $T, S, U$,  written in terms of the physical gauge boson vacuum polarizations as \cite{TSU2}
\begin{eqnarray}
\label{STU2}
T & = & \frac{4 \pi }{e^2 c_W^2 m_Z^2} \left[ A_{WW}(0)- c_W^2 A_{ZZ} (0) \right] \,, \nn \\[3mm]\nn
S & = &16 \pi \frac{ s_W^2 c_W^2}{e^2}  \left[  \frac{A_{ZZ}(m_Z^2)- A_{ZZ}(0) }{m_Z^2}- A'_{\gamma \gamma}(0) -\frac{ (c_W^2-s_W^2) }{c_W s_W} A_{\gamma Z}'(0) \right] \,, \\[3mm]\nn
U &=& -16 \pi \frac{s_W^2}{e^2} \Biggl[ \dfrac{A_{WW}(m_W^2)- A_{WW}(0) }{m_W^2}- c_W^2 \frac{A_{ZZ}(m_Z^2)- A_{ZZ}(0) }{m_Z^2}+ \\[3mm] && -s_W^2 A_{\gamma\gamma}' (0) - 2 s_W c_W A_{\gamma Z}' (0) \Biggr]\,,
\end{eqnarray}
where  $s_W,c_W$ are sine and cosine of $\theta_W$ and $e$ is the electric charge. Roughly speaking  for any $SU(2)_L$ doublet  $T,S,U$  are sensitive to the mass splitting of the doublet  components and thus vanish in the limit of degenerate masses \cite{Drees:1990dx}. As it can be easily checked by looking at the spectra shown in figure \ref{figure:spectra} the SUSY breaking scale is so high that the components of the $SU(2)_L$ doublets happen to be still degenerate after the EW spontaneous symmetry breaking, and right and left sfermion mixing is thus completely negligible.  Contemporaneously at the EW scale the wino and  the light chargino form a  degenerate doublet, so the EW parameters do not receive any new contribution arising from new particles, the unique contribution being that of the SM-like Higgs.  The latter, for a  mass around 135 GeV, is in perfect agreement  with the data \cite{PDG}.

\subsection{Flavor constraints}

A detailed analysis of flavor processes is beyond the purposes of this work. Intuitively such processes should not further constrain the model because of the very heavy  sparticle spectrum, but we should take care of them because of the yukawa structure of the SUSY breaking mediation mechanism. Being the third family the heaviest and the first two almost degenerate and lighter we are in presence of a  hierarchical squark spectrum inverted  with respect to  the discussion developed in \cite{Giudice:2008uk}. However we may use their formalism to estimate the contribution to $\Delta F=1,2$ processes. 

As an example we may consider the gluino loop contributions in the down sector. The latter in the case of $d_i^L\to d_j^L$ ($\Delta F=1$)  and $d_i^L\leftrightarrow d_j^L$ ($\Delta F=2$)   may be parametrized as
\begin{eqnarray}
\label{deltaF}
A(\Delta F=1)&=&\mathcal{W}_{d_i^L \tilde{D}_I}\,f\left( \frac{m^2_{\tilde{D}_I}}{M^2_3} \right)\, \mathcal{W}^*_{d_j^L \tilde{D}_I} \nn\\
A(\Delta F=2)&=&\mathcal{W}_{d_i^L \tilde{D}_I}\,\mathcal{W}_{d_i^L \tilde{D}_J} g \left( \frac{m^2_{\tilde{D}_I}}{M^2_3} ,\frac{m^2_{\tilde{D}_J}}{M^2_3} \right)\, \mathcal{W}^*_{d_j^L \tilde{D}_I} \,  \mathcal{W}^*_{d_j^L \tilde{D}_J}\,,\nn
\end{eqnarray}
where $f$ and $g$ are loop functions, in particular  $g(x,y)= g(x)-g(y)/(x-y)$. $\mathcal{W}$ diagonalizes the full $6\times 6$ down squark mass matrix in the basis in which the down quarks are diagonal, namely ${\mathcal{M}}^2_D$. In our specific framework $\mathcal{W}$ is very simple and  block diagonal because  the soft terms are so heavy that  we can neglect left-right squark mixing. In particular this means that gluino loops cannot mediate $\Delta F=1$ processes, like $b\to s \gamma$, and in the following we will concentrate only on $\Delta F=2$ processes.  If we assume that $Y_u$ and $Y_d$ have a Froggatt Nielsen symmetric structure \cite{FroggattNielsen} -- thus implying that $V_{CKM}\sim V_L^u\sim V_L^d\sim V_R^u\sim V_R^d $ -- the structure of $\mathcal{M}^2_{D_{L,R}}$ at $M$ in our specific framework is roughly given by
\begin{eqnarray}
\label{squarkmat}
\mathcal{M}^2_{D_{L,R}}=  \Biggl[Lp^2 B^{r}\mathcal{Y}_{r} h_0^2 + Lp^3 \Bigl( C^{i,j,k}_1 r^2_i r^2_j r^2_k + C^{i,j,k}_2 r^2_i r^2_j \mathcal{Y}_k + C^{i,j,k}_3 r^2_i \mathcal{Y}_j \mathcal{Y}_k \Bigr)\Biggr] B_{\phi}^2
\end{eqnarray}  
where the notation is the same of equations (\ref{2loopcoeff}) and (\ref{3loopcoeff}). Equation (\ref{squarkmat}) may be re-written as

\begin{eqnarray}
\label{squarkmat}
\mathcal{M}^2_{D_{L,R}}=  \frac{M_3^2}{Lp^2}\left[ \mathcal{Y} \frac{h_0^2}{g_3^4 h_t^4} + Lp \left(C^{i,j,k}_1 \frac{r^2_i r^2_j r^2_k}{g_3^4 h_t^4} + C^{i,j,k}_2 \frac{r^2_i r^2_j}{g_3^4 h_t^4} \mathcal{Y}_k + C^{i,j,k}_3 \frac{r^2_i}{g_3^4 h_t^4} \mathcal{Y}_j \mathcal{Y}_k \right) \right],
\end{eqnarray} 
where $\mathcal{Y} = B^r \mathcal{Y}_r$. Among the terms that arise at three loop level the dominant one is that proportional to $C_1$. 
The  diagonal two loop entries dominate over the three loop ones if 
\begin{equation}
{(\mathcal{Y})_{ii}} > Lp \, C^{i,j,k}_1 \frac{r^2_i r^2_j r^2_k}{h_0^2} \sim Lp \, c_g \, g^4 \sim 10^{-1}\,,
\end{equation}
where $c_g \sim \mathcal{O}(10)$, that clearly is realized only for the third family. Thus the two loop term controls the heavy third familiy squark masses and the off diagonal entries, while the first three loop term dominates the degenerate two lightest families. For the following discussion we are interested in the ratios 
\begin{eqnarray}
\label{ratios}
\Delta^{L,R}_{12}&=&\frac{(\mathcal{M}^2_{D_{L,R}})_{12}}{(\mathcal{M}^2_{D_{L,R}})_{22}}\sim \frac{1}{c_g Lp}\frac{(\mathcal{Y})_{12}}{g^4}\sim  10 \lambda_C^5 \frac{m_b^2}{m_t^2}\frac{\tan\beta^2}{g^4}\sim 10^{-3} \div 10^{-2}\,,\nn\\
&&\nn\\
\Delta^{L,R}_{23}&=&\frac{(\mathcal{M}^2_{D_{L,R}})_{23}}{(\mathcal{M}^2_{D_{L,R}})_{33}}\sim \frac{(\mathcal{Y})_{23}}{(\mathcal{Y})_{33}}\sim \lambda_C^2 \sim 10^{-2} \,,\nn\\
&&\nn\\
\Delta^{L,R}_{13}&=&\frac{(\mathcal{M}^2_{D_{L,R}})_{13}}{(\mathcal{M}^2_{D_{L,R}})_{33}}\sim \frac{(\mathcal{Y})_{13}}{(\mathcal{Y})_{33}}\sim  \lambda_C^3 \sim 10^{-3} \,.
\end{eqnarray}
In equation (\ref{ratios}) $m_t$ ($m_b$) are the top (bottom) quark mass, $\lambda_C$ the Cabibbo angle $\sim 0.2$ and as usual $\tan\beta=v_u/v_d$. According to our considerations $\mathcal{W}$ has the structure
\begin{equation}
\mathcal{W}= \left(\begin{array}{cc} \mathcal{W}^L&0\\0&
 \mathcal{W}^R \end{array} \right)\,.
\end{equation}
with
\begin{eqnarray}
\mathcal{W}_{{L,R}}&\simeq&  \left( \begin{array}{ccc} \cos \theta^L_{12}&\sin \theta^L_{12}&\Delta^L_{13}\\ -\sin^L \theta_{12}&\cos \theta^L_{12} &\Delta^L_{23}\\ 
\Delta^L_{23}\sin\theta^L_{12} -\Delta^L_{13}\cos \theta^L_{12} &-\Delta^L_{23}\cos\theta^L_{12} -\Delta^L_{13}\sin \theta^L_{12} &1\end{array}\right)\,,
\end{eqnarray}  
where
\begin{equation}
\tan 2 \theta_{12}=2 \frac{(\mathcal{M}_{D_{L}})_{22}\Delta^L_{12}}{(\mathcal{M}_{D_{L}})_{22}-(\mathcal{M}_{D_{L}})_{11}}\sim \frac{(\mathcal{Y})_{12}}{(\mathcal{Y})_{22}}\,.
\end{equation}
The generalization to up squarks and charged sleptons is trivial.

Following \cite{Giudice:2008uk} in our model we have
\begin{eqnarray}
\label{deltaF2b}
A(\Delta F=2)&\sim&g^{(1)}(x)(\hat{ \delta}^{LL}_{ij})^2+ \frac{ x^2}{3!} g^{(3)}(x)(\delta_{ij}^{LL})^2\,,
\end{eqnarray}
with $x={\tilde{m}^2_{L}}/{M_3^2}$ and  we can set $\tilde{m}^2_{\tilde{d}_{L,R}} \sim  \tilde{m}^2_{\tilde{s}_{L,R}} \sim \tilde{m}^2_{{L,R}}$ because of their approximate degeneracy.  The $\delta$ parameters are defined as
\begin{eqnarray}
\label{deltas}
\hat{ \delta}^{LL}_{ij} &=& \mathcal{W}^L_{d_i \tilde{d}} \mathcal{W}^{L*}_{d_j \tilde{d}}+ \mathcal{W}^L_{d_i \tilde{s}} \mathcal{W}^{L*}_{d_j \tilde{s}}=\delta_{ij}- \mathcal{W}^L_{d_i \tilde{b}} \mathcal{W}^{L*}_{d_j \tilde{b}}\nn\\
&\simeq& \delta_{ij}-\delta_{33}-\left[(\delta_{ik}\delta_{j3}+\delta_{jk}\delta_{i3})\Delta_{k3}^L\right]_{k=1,2}\,,\nn\\
{ \delta}^{LL}_{ij} &=& (\delta_{i2}\delta_{j3}+\delta_{j2}\delta_{i3})  \Delta^L_{12} \,.
\end{eqnarray}
Clearly the term that depends  on $\hat{ \delta}^{LL}_{ij}$ controls $B_{d,s}-\overline{B}_{d,s}$ oscillation, while that on ${ \delta}^{LL}_{ij}$ controls $K-\overline{K}$. 
In the limit $x \gg 1$ it turns out that \cite{Buras:1997ij} $g^{(1)}(x),g^{(3)}(x) \sim 1/x,1/x^3$ respectively, and we get
\begin{equation}
\label{deltaF2b}
A(\Delta F=2)=\mathcal{F}^{LL}_{ij}\sim\alpha_s C_q \left(\frac{M_3^2}{\tilde{m}^2}\right)\left[ (\hat{ \delta}^{LL}_{ij})^2+\frac{1}{6} ({ \delta}^{LL}_{ij})^2\right]   \,,
\end{equation}
where $C_q$ is a color factor. From \cite{Atwood:1996vj} we have
\begin{equation}
\label{meson}
\Delta M_F= M_F f_F^2 B_F \frac{8}{3M_3^2}\mathcal{F}^{LL}_{ij}\,,
\end{equation}
where $B_F$ is  a parameter of order 1, $f_F$ is the decay constant of the meson $F = K, B_d, B_s$ and $ij$ the transition responsible of its oscillations.  By combining equations (\ref{deltas}), (\ref{deltaF2b}) and (\ref{meson}) we get
\begin{eqnarray}
\label{res}
\Delta M_K&=& C_q \alpha_s^2 M_K  f_K^2B_K \frac{4}{9 \tilde{m}^2}(\Delta_{12}^L)^2  \sim M_K \left( \frac{f_K^2}{\mbox{GeV}^2} \right) B_K(10^{-14} \div 10^{-16})\,,\nn\\
&&\nn\\
\Delta M_{B_d}&=& C_q\alpha_s^2M_{B_d} f_{B_d}^2B_{B_d} \frac{8}{3 \tilde{m}^2}(\Delta_{13}^L)^2  \sim M_{B_d} \left( \frac{f_{B_d}^2}{\mbox{GeV}^2} \right) B_{B_d} 10^{-15}\,.\nn\\
&&\nn\\
\Delta M_{B_s}&=&C_q \alpha_s^2 M_{B_s} f_{B_s}^2B_{B_s} \frac{8}{3 \tilde{m}^2}(\Delta_{23}^L)^2  \sim M_{B_s} \left( \frac{f_{B_s}^2}{\mbox{GeV}^2} \right) B_{B_s} 10^{-13}\,.\nn
\end{eqnarray}
By comparing our  results and the experimental bounds as reported in table \ref{table1} we see  that the flavor processes mediated by the sfermions are a few orders of magnitude below the experimental bounds. Similar results are expected in the case of loops in which circulate other superpartners.

We also briefly comment about the possibility of gravity mediated FC processes. The latter are generated by the complete democracy of gravity mediated interactions in flavour space and thus can be neglected only if they are subleading with respect to the other contributions. Such a constraint imposes bounds on the mass of the gravitino that will be taken into account in Section \ref{cosmology}.

\begin{table}[!h]
\begin{center}
\begin{tabular}{|c||c|c|c|c|c|c|}
\hline
$B_F$ & $M_F$ (GeV) & $f_F$ (GeV) & $B_F$ & $|\Delta M^{exp}_F|$  (GeV)&$|\Delta M^{res}_F|$  (GeV) \\
\hline
$B_d$ & $5.2795$ & $0.1928 \pm 0.0099$ & $1.26 \pm 0.11 $ & $(3.337 \pm 0.006 ) \times 10^{-13}$&$<10^{-16}$ \\
$B_s$ & $5.3664$ & $0.2388 \pm 0.0095$ & $1.33 \pm 0.06 $ & $(1.170 \pm 0.008 ) \times 10^{-11}$&$ <10 ^{-14}$ \\
$K$ & $0.497614$ & $0.1558 \pm 0.0017$ & $0.725 \pm 0.026 $ & $(3.500 \pm 0.006 ) \times 10^{-11}$& $<10 ^{-16} $ \\
\hline
\end{tabular}
\end{center}
\caption{\it Properties of neutral mesons \cite{Buras:2010mh} and the model predictions. The last column report  our  rough estimation of the meson oscillation mass splitting  $\Delta M^{res}_F$ taking as reference values the spectrum given in figure \ref{figure:spectra}.}
\label{table1}
\end{table}

\subsection{Cosmology}
\label{cosmology}

In minimal gauge mediation the LSP is usually the gravitino \cite{Giudice:1998bp}. Here we discuss if our framework behaves as the minimal case or not, allowing  the bino or the wino to be the LSP.  

The gravitino takes mass through gravitational interactions: their coupling strenght is given by the inverse reduced Planck mass and, assuming a vanishing cosmological constant, the mass $m_{3/2}$ is
\begin{equation}
m_{3/2} = \frac{F_0}{\sqrt{3} M_{P}}\,,
\end{equation}
where $M_{P} = (8 \pi G_N)^{-1/2}$ is the reduced Planck mass, $F_0$ is the total contribution of the F-term SUSY breaking vev to the vacuum energy, thus $V = F_0^2$ in the minimum. Actually the effective F-term vev  felt by the messenger $\hat{\Phi}$ is related to $F_0$ by means of the superpotential interaction $k \hat{X}\, \hat{\Phi}\,\hat{ \Phi}$, thus being $F_0 = F / k$. The ratio among the two quantities directly reflects the way in which SUSY breaking is mediated, and in our case  it is simply given by the coupling constant $k\lesssim1$  to preserve perturbativity at high energy scale. The gravitino mass can thus be rewritten as 
\begin{equation}
\label{mgr}
m_{3/2} = \frac{B_\phi M}{k \sqrt{3} M_{P}}.
\end{equation}
To point out the nature of the LSP it turns useful rewriting the gravitino mass in terms of the wino one. Reminding  that 
\begin{equation}
M_2 = Lp^2 g_2^2 h_0^2 B_\phi \, , 
\end{equation}
we get
\begin{equation}
\label{mgrM2}
m_{3/2} = \Bigl( \frac{M_2(M)}{250 \GeV} \Bigr) \Bigl( \frac{0.7}{g_2(M)} \Bigr)^2 \frac{1}{h_0^2(M)} \frac{0.9}{k(M)} 3.4 \times 10^{-12} M,
\end{equation}
where $M$ is the $\hat{\Phi}$ superfield mass scale and thus gives rise to the SUSY breaking boundary conditions. As one can easily see from the formula above the gravitino is the LSP if a relatively low energy SUSY breaking takes place. If the scale of SUSY breaking is of order $M \gtrsim 10^{14} \GeV$ than the relative weight of the adimensional parameter entering into equation (\ref{mgrM2})  establishes whether the LSP is the gravitino or neutralino, while for larger values of $M$, such as the GUT scale, the gravitino is surely not the LSP of the framework. At first let us briefly comment on this possibility.

lf R-parity is conserved and the neutralino (wino or bino) is the LSP of the model the decay of the gravitino must happen before Big Bang Nucleosynthesis (BBN) in order not to destroy the successful predictions of BBN itself \cite{BBN}. In particular if gravitino decays have to be completed before BBN at $t \sim 1 s$ we must have 
\begin{equation}
m_{3/2} \gtrsim 10 \TeV \Longrightarrow M \gtrsim 10^{15} \div 10^{16} \GeV\,,\quad B_\phi \lesssim 10^{7} \div 10^{8} \GeV\,.
\end{equation} 
Such a large gravitino mass would give rise to a very large contribution to sfermion mass matrices that would induce dangerous FCNC and our hypothesis to neglect gravitino contribution to sfermion masses would fail. 

Consequently in our scenario the gravitino has to be the LSP, condition that can be rephrased as $M \lesssim 10^{14}$ GeV. On the other side bounds on gluino mass impose $B_\phi >10^6$ GeV and $M > 10^7$ GeV implying  $m_{3/2} > 1\keV$.  Light gravitino DM scenarios are therefore not realizable in our framework, while the possibility of  a superWIMP DM  is left open \cite{feng}. The analysis of the gravitino as superWIMP candidate is left for further studies.

\section{Outlook}
\label{outlook}

In section \ref{cosmology} we deduced that the allowed range for  $M$  is $10^7 \GeV < M \lesssim 10^{14} \GeV$  in order not to affect FCNC processes, gauge coupling unification and gluino mass bounds. One very interesting possibility is promoting $\hat{\Phi}$ to be both the SUSY breaking messenger and the source of light neutrino masses, through its fermionic component. This picture shares features both with the $\nu$GMSB model  \cite{FileviezPerez:2009im} and  the  bilinear  R-parity breaking scenarios \cite{Hirsch:2000ef,Ross:1984yg}, since $\hat{\Phi}$ is a singlet superfield.  As in  \cite{FileviezPerez:2009im}   light neutrinos receive type I seesaw-like mass contribution arising from the  explicit R-parity breaking term -- the coupling $y_i\hat{L}_i \hat{H}_u \hat{\Phi}$ -- yielding
\begin{equation}
\label{co1}
m_\nu^I \sim \frac{v_u^2}{M} y_i\cdot y_i^T\,,
\end{equation}
where the $y_i$ are thought as three dimensional vector columns. When sneutrinos  and the scalar component of $\hat{\Phi}$  acquire a  tiny vev a further contribution to left handed neutrino masses is generated. This contribution is nothing but that presents in models with explicit linear R-parity breaking term and discussed in details in  \cite{Hirsch:2000ef}. Using their notation and  assuming a spectrum similar to that given in figure \ref{figure:spectra} the second contribution to neutrino masses is given by 
\begin{equation}
m_\nu^{Rbr} \sim \frac{g_2^2+ 3/5 g_1^2}{4 \mu^2 M_0} \Lambda_i\cdot \Lambda_i^T\,,
\end{equation}
where we have approximated $M_1\sim M_2\sim M_0 \ll \mu$ and $\Lambda_i$ is defined as
\begin{equation}
\Lambda_i= \mu v_i + y_i v_d v_\phi\,,
\end{equation}
with $v_i$ ($v_\phi$) is the sneutrino ($\Phi$) vev. As long as $v_i$ and $y_i$ are disaligned\footnote{This may be realized because of the sneutrino  soft mass terms that are not aligned to $y_i y_i^T$. Indeed the soft sneutrino masses receive a contribution both proportional to $y_i$ from  $y_i\hat{L}_i \hat{H}_u \hat{\Phi}$ and to $ Y_e^\dagger Y_e$ from $(Y_e)_{ij}\hat{L}_i \hat{H}_d \hat{E}_j $ . },  neglecting the one loop contributions, the effective light neutrino mass matrix has two non vanishing eigenvalues and lepton mixing is completely determined. 

This scenario is quite appealing for its predictivity in neutrino sector and we leave a detailed analysis to a future project \cite{inprogress}. Notice that in this case  a late decaying gravitino should be the  DM candidate \cite{Restrepo:2011rj}. 

\section{Summary and conclusions}
\label{conc}

In this paper we focused on the idea of yukawa-gauge mediation in what we may define its minimal version. The minimality resides in the small number of extra degrees of freedom with respect to those of the MSSM. The chiral superfield $\hat{\Phi}$  is the first messenger of SUSY breaking and due to its singlet nature SUSY breaking happen to be communicated to the MSSM fields in two sequential steps, after the subsequent coupling of a charged superfield sector. Such second messengers are identified with the MSSM Higgs fields with the addition of an extra color triplets that ensure a large enough gluino mass. This framework gives rise to a novel split spectrum that is consistent with the indication arising from the most recent data. The peculiarities of such a spectrum are an inverted hierarchy among the sfermions of the third family and those of the first two and an almost complete pureness of the gaugino components in the lightest neutralinos and chargino. The mass scale of the scalar superpartners is fairly heavy, above few TeVs, and only the gauginos, that are longly lived because of a quite large gravitino mass, will be soon in the range of LHC searches. We did not analyze the specific signatures of the model presented, however  from the shape of the predicted spectrum we may easily  deduce  that few processes could be testable at LHC. In particular the preferred channel seems to be an excess of events over the SM background in the Mono-Jet channel, owed to ISR combined with the production of two gluinos. As far as the EWSB is concerned we computed the one loop effective potential and showed that a single SM-like Higgs of mass $130 \div 135 \GeV$ is expected. 

In the scenario gauge coupling unification is still achievable and one could easily think to extend the analysis to an SU(5) description of the framework. Thanks to the peculiar structure of the soft terms the model easily satisfies the constraints imposed by SM precision measurements, as EWPT and FCNC. We estimated the one loop contribution to $\Delta F=2$ meson oscillations mediated by the gluinos and we have shown that is extremely suppressed and far from experimental bounds.

To prevent large democratic contribution to sfermion masses, possibly mediating FCNCs, and not to affect BBN the gravitino has to be the LSP of our scenario. Combining the request for a gravitino LSP with the bounds on gluino masses gives us the range in which the SUSY breaking scale $M$ has to lie, namely $10^7< M \leq 10^{14}$. 

Finally we sketched how a R-parity breaking version of our scenario could be deeply interesting in neutrino phenomenology and compatible with a late decaying gravitino as DM candidate. We leave the full analysis to \cite{inprogress}.

\section{Acknowledgments}
The authors would like to thank G. Arcadi, M. Lattanzi and M. Pierini for useful discussions.

\appendix

\section{Gauge coupling unification}
\label{appendix:gaugeunification}

In this appendix we discuss the details of gauge coupling evolution, whose results are reported in section \ref{section:gaugeunification}; in particular we show that unification is easily realized in our framework. 

The one loop evolution of the gauge couplings is given by 
\begin{equation}
(4 \pi)^2 \frac{d g_i}{dt} = g_i^3 b_i  \,,
\end{equation}
where the $b_i$'s have to be calculated considering all the fields that are charged under the i-th interaction. For a generic theory the different contributions owed to scalars and fermions may be obtained  as in \cite{2loopgeneral}. Thus we calculate the various $b_i$'s considering only those fields present in the effective field theory at a given scale: indeed at scale $\mu$ all the fields heavier than $\mu$ decouple and do not contribute to the running. 

Our framework is characterized by the presence of five scales: 
\begin{itemize}
\item [-] winos and bino, $M^{(4)}_{SUSY} \sim 100 \div 300 \GeV$;
\item[-] gluinos, $M^{(3)}_{SUSY} \sim 1 \div 1.5 \TeV$;
\item [-] light sfermions, $M^{(2)}_{SUSY} \sim 4 \div 7 \TeV$;
\item [-] heavy sfermions, heavy Higgs and higgsinos, $M^{(1)}_{SUSY} \sim 30 \div 50 \TeV$;
\item [-] extra triplets, $M_T \sim 10^{14} \div 10^{15} \GeV$.
\end{itemize}
The different $b_i$'s at the energy scale $\mu$ are: 
\begin{itemize}
\item $\mu <M^{(4)}_{SUSY}$: the theory coincides with the SM thus $b_i = \left\{ \frac{41}{10}, -\frac{19}{6}, -7 \right\}$;
\item $M^{(3)}_{SUSY} < \mu < M^{(4)}_{SUSY}$:  the theory is SM+ winos + bino thus $b_i = \left\{ \frac{41}{10}, -\frac{11}{6}, -7 \right\}$;
\item $M^{(2)}_{SUSY} < \mu < M^{(3)}_{SUSY}$: the theory is SM + winos + bino + gluino thus $b_i = \left\{ \frac{41}{10}, -\frac{11}{6}, -5 \right\}$;
\item $M^{(2)}_{SUSY} < \mu < M^{(1)}_{SUSY}$: the theory is SM+ winos + bino + gluino + first two families sfermions thus $b_i = \left\{ \frac{163}{30}, -\frac{1}{2}, -\frac{11}{3} \right\}$;
\item $M^{(1)}_{SUSY} < \mu < M_T$, the theory is MSSM thus $b_i = \left\{ \frac{33}{5}, 1, -3 \right\}$;
\item $\mu > M_T$, the theory is MSSM + heavy triplets thus $b_i = \left\{ 7, 1, -2 \right\}$. 
\end{itemize}
We can check the compatibility of gauge unification with the observables at the EW scale as follows. The  low energy values at $M_Z$, the Z boson mass scale, of $\alpha$, $\sin^2 \theta_W$ and $\alpha_s$  -- in the $\overline{MS}$ renormalization scheme -- are given by \cite{PDG} 
\begin{eqnarray}
M_Z = 91.1876 \pm 0.0021 \GeV \,, \nn\\
\alpha(M_Z)^{-1} = 127.916 \pm 0.015 \,, \nn\\ 
\sin^2 \theta_W (M_Z) = 0.23150 \pm 0.00016 \,, \nn\\
\alpha_s(M_Z) = 0.1184 \pm 0.0012 \nn\,.
\end{eqnarray}
The observable with the biggest uncertainties is $\alpha_s$. We evolve  $g_1$ and $g_2$ to high energies and we determine a tentative unification scale. Then we run back the strong coupling and check if it  is compatible with the experimental value. We find that for the typical scales corresponding to the spectrum shown in figure \ref{figure:spectra}  the strong coupling calculated with this procedure falls always within three sigmas of the experimental value. 

A brief  comment about two loop evolution. The two loop RGEs \cite{2loop} yield non negligible contributions that have to be considered when discussing precision observables. In particular when going beyond one loop it is important not only to use two loop RG equations, but also consider the one loop threshold corrections in order to be consistent. Keeping into account the two contributions it can be seen that there is a partial cancellation between them, thus the net contribution tends to  improve the agreement with the best fit value of $\alpha_s$, and gauge coupling unification is preserved.

\section{RG evolution of the effective theory}
\label{appendix:higgs}

In section \ref{higgs} we have seen that below $M^{(1)}_{SUSY}$ the heavy fields start to decouple from the theory. In the following we  write down the evolution equations for the  parameters involved  to get the Higgs running mass at the minimum of the potential. As already anticipated in appendix \ref{appendix:gaugeunification} we spotted the presence of four different intermediate scales to consider. Not to be redundant we do not re-write the gauge coupling RGEs, that can be found in appendix \ref{appendix:gaugeunification}. 

\subsection*{Between $M_{SUSY}^{(1)}$ and $M_{SUSY}^{(2)}$}

In this region the heavy third family sfermions, the Higgsinos and the heavy Higgs doublet have decoupled. The relevant equations needed for the evolution of the various parameters are the following.  The couplings evolve through
\begin{equation}
(4 \pi)^2 \frac{d \xi}{dt} = \beta_{\xi}.
\end{equation}
We considered the various $\beta_\xi$ and $\gamma_{h}$, the anomalous dimensions, in the third family approximation, where only the top yukawa coupling is relevant:
\begin{eqnarray}
\beta_{M_1} &=& 8 g_1^2 M_1 \,,\\
\beta_{M_2} &=& - 4 g_2^2 M_2 \,,\\
\beta_{M_3} &=& - 10 g_3^2 M_3 \,,\\
\beta_{Y_t} &=& Y_t \left( \frac{9}{2} Y_t^2 - \frac{17}{20} g_1^2 - \frac{9}{4} g_2^2 - 8 g_3^2 \right)\,, \\
\beta_{\lambda} &=& 12 \lambda^2 + \lambda \left( 12 Y_t^2 - \frac{9}{5} g_1^2 - 9 g_2^2 \right) + \frac{9}{2} \left( \frac{3}{50} g_1^4 + \frac{g_2^4}{2} + \frac{g_1^2 g_2^2}{5} \right) - 12 Y_t^4 \,,\\
\beta_{m_h^2} &=& m_h^2 ( 6 \lambda + 6 Y_t^2 - \frac{9}{2} g_2^2 - \frac{9}{10} g_1^2 ) \,,\\
\gamma_{h} &=& 3 Y_t^2 - \frac{9}{4} g_2^2 - \frac{9}{20} g_1^2 .  
\end{eqnarray}

\subsection*{Between $M_{SUSY}^{(2)}$ and $M_{SUSY}^{(3)}$}

At scale $M_{SUSY}^{(3)}$ the light sfermions (namely those of the first two families) decouple. The various $\beta_\xi$ and $\gamma_{h}$ in the third family approximation are now:
\begin{eqnarray}
\beta_{M_1} &=& 0 \,,\\
\beta_{M_2} &=& - 12 g_2^2 M_2 \,,\\
\beta_{M_3} &=& - 18 g_3^2 M_3 \,,\\
\beta_{Y_t} &=& Y_t \left( \frac{9}{2} Y_t^2 - \frac{17}{20} g_1^2 - \frac{9}{4} g_2^2 - 8 g_3^2 \right) \,,\\
\beta_{\lambda} &=& 12 \lambda^2 + \lambda \left( 12 Y_t^2 - \frac{9}{5} g_1^2 - 9 g_2^2 \right) + \frac{9}{2} \left( \frac{3}{50} g_1^4 + \frac{g_2^4}{2} + \frac{g_1^2 g_2^2}{5} \right) - 12 Y_t^4 \,,\\
\beta_{m_h^2} &=& m^2 ( 6 \lambda + 6 Y_t^2 - \frac{9}{2} g_2^2 - \frac{9}{10} g_1^2 ) \,,\\
\gamma_{h} &=& 3 Y_t^2 - \frac{9}{4} g_2^2 - \frac{9}{20} g_1^2 .  
\end{eqnarray}

\subsection*{Between $M_{SUSY}^{(3)}$ and $M_{SUSY}^{(4)}$}

Below $M_{SUSY}^{(3)}$ also the gluinos cease to be part of the theory. $\beta_\xi$ and $\gamma_{h}$ are now given by
\begin{eqnarray}
\beta_{M_1} &=& 0 \,,\\
\beta_{M_2} &=& - 12 g_2^2 M_2 \,,\\
\beta_{Y_t} &=& Y_t \left( \frac{9}{2} Y_t^2 - \frac{17}{20} g_1^2 - \frac{9}{4} g_2^2 - 8 g_3^2 \right) \,,\\
\beta_{\lambda} &=& 12 \lambda^2 + \lambda \left( 12 Y_t^2 - \frac{9}{5} g_1^2 - 9 g_2^2 \right) + \frac{9}{2} \left( \frac{3}{50} g_1^4 + \frac{g_2^4}{2} + \frac{g_1^2 g_2^2}{5} \right) - 12 Y_t^4 \,,\\
\beta_{m_h^2} &=& m_h^2 ( 6 \lambda + 6 Y_t^2 - \frac{9}{2} g_2^2 - \frac{9}{10} g_1^2 ) \,,\\
\gamma_{h} &=& 3 Y_t^2 - \frac{9}{4} g_2^2 - \frac{9}{20} g_1^2 .  
\end{eqnarray}

\subsection*{Below $M_{SUSY}^{(4)}$}

Below $M_{SUSY}^{(4)}$ the theory is nothing but the SM. The evolution can be obtained as in \cite{2loopSM}. For the sake of completeness we report the relevant equations.  The relevant $\beta_\xi$'s  are
\begin{eqnarray}
\beta_{Y_t} &=& Y_t \left( \frac{9}{2} Y_t^2 - \frac{17}{20} g_1^2 - \frac{9}{4} g_2^2 - 8 g_3^2 \right) \,,\\
\beta_{\lambda} &=& 12 \lambda^2 + \lambda \left( 12 Y_t^2 - \frac{9}{5} g_1^2 - 9 g_2^2 \right) + \frac{9}{2} \left( \frac{3}{50} g_1^4 + \frac{g_2^4}{2} + \frac{g_1^2 g_2^2}{5} \right) - 12 Y_t^4 \,,\\
\beta_{m_h^2} &=& m_h^2 ( 6 \lambda + 6 Y_t^2 - \frac{9}{2} g_2^2 - \frac{9}{10} g_1^2 ) \,,\\
\gamma_{h} &=& 3 Y_t^2 - \frac{9}{4} g_2^2 - \frac{9}{20} g_1^2 .  
\end{eqnarray}

\section{Calculation of the SUSY breaking terms}

In this appendix we derive the SUSY breaking terms in our theory. In general such terms can be derived by means of diagrammatic loop calculation in which  the SUSY breaking F-term vev enters. In our scenario such a calculation  is quite involved, as it would require to consider graphs up to three loops. It is far more convenient to use an approach based on the properties of SUSY theories renormalization. Following the seminal paper \cite{Giudice:1997ni} we review this method and its realization in our framework. 

\subsection*{General theory}

As already discussed in section \ref{model} we parametrize the breaking of SUSY through the presence of a single chiral superfield $\hat{X}$ taking vev both in its scalar and auxiliary components, $\langle \hat{X} \rangle = M/k + \theta^2 F/k$, where k is a coupling constant that has been reabsorbed in the vevs for later convenience. The only source for the appearance of soft terms is the vev $F$. Thus the SUSY breaking contributions can be casted in an expansion in terms of powers of $F$, or to be more precise, in terms of the dimensionless parameter $F/M^2$. If interested in the regime $F \ll M^2$, one can keep track of the SUSY breaking effects in a manifestly supersymmetric framework, considering soft terms just as small modifications of the latter. Being a bit sloppy we can explain the procedure as follows. Let us consider a SUSY gauge theory based on the gauge group $\mathcal{G}$. It is well known that in SUSY theories the renormalization effects are owed to the renormalization of the kinetic terms of gauge and matter, while the superpotential does not renormalize. Thus one can calculate the evolution of the matter wave functions and of the gauge couplings in the SUSY limit from the high cutoff scale $\Lambda_{UV}$ down to low energies across the scale $M$ using the RGEs and then substitute $M \rightarrow \sqrt{X X^\dagger}$ into the gauge real couplings $\mathcal{R}(M,\mu)$\footnote{The real coupling is the gauge coupling defined for canonically normalized fields and it is different from the holomorphic coupling. The relation among the two is given in equation (\ref{real-holomorphic}). The reason for the presence of two different couplings is well explained in \cite{ArkaniHamedMurayama}. In particular while the holomorphic coupling has the nice property of renormalizing just at one loop, the real coupling is the canonically normalized one that actually couples to matter, thus giving the strenght of the interaction.}  and in the wave function renormalizations $Z_r(M,\mu)$. Since the $\hat{X}$ superfield takes  both scalar and F-term vevs, such a procedure implies the appearance of soft terms in the lagrangian of the theory. Extracting them is then simply a matter of knowing the $F$-dependence of the various $\mathcal{R}$ and $Z_r$'s.

To be definite we start by considering the high energy theory defined by the lagrangian
\begin{eqnarray}
\label{lagrangianUV} 
\mathcal{L} &= & \int d^4\theta \Biggl[ Z^>_M \hat{\Phi}^\dagger e^{V(\hat{\Phi})}  \hat{\Phi} + \sum_r Z^>_r \hat{Q}_r^\dagger e^{V(\hat{Q}_r)} \hat{Q}_r \Biggr] + \nonumber\\
&& + \int d^2\theta \frac{1}{2}S^> \tr\Bigl(W^\alpha W_\alpha\Bigr) + h.c. \nonumber\\
&& + \int d^2\theta (\lambda X \hat{\Phi} \hat{\Phi} +\mu_{rs}\hat{Q}_r \hat{Q}_s+ \lambda_{rst} \hat{Q}_r \hat{Q}_s \hat{Q}_t )+ h.c.
\end{eqnarray}
where $\hat{\Phi}$ is a chiral messenger (to be general we assume it charged under $\mathcal{G}$) and the  $\hat{Q}_r$ are the matter superfields, $\hat{X}$ is the chiral superfield taking vevs and $\mu_{rs},\lambda_{rst}$ are the SUSY superpotential mass terms and yukawa interactions respectively. Finally $Z_M^>, Z^>_r,S^>$ are the renormalization wave functions above the scale $M$. At such a scale the messenger superfield $\hat{\Phi}$ takes mass and decouples, thus at lower energies one gets 
\begin{eqnarray}
\label{lagrangianlowen}
\mathcal{L} &=& \int d^4\theta  \sum_r Z^<_r \hat{Q}_r^\dagger e^{V(\hat{Q}_r)} \hat{Q}_r  + \int d^2\theta\frac{1}{2} S^< \tr\Bigl(W^\alpha W_\alpha\Bigr) + h.c. \nonumber\\
&& +\int d^2\theta(\mu_{rs}\hat{Q}_r \hat{Q}_s+ \lambda_{rst} \hat{Q}_r \hat{Q}_s \hat{Q}_t )+ h.c.\,.
\end{eqnarray}
The above equations (\ref{lagrangianUV}) and (\ref{lagrangianlowen}) define the theory in which the RG evolution takes place. Once one knows the $\gamma$'s of the superfields and the $\beta$ functions of the couplings involved it is possible to determine the $\mathcal{R}(M,\mu)$'s and the $Z_r(M,\mu)$'s ($Z_r$ is defined as $Z_r^>$ above $M$ and $Z_r^<$ below it). By substituting  $M \rightarrow \sqrt{X X^\dagger}$ and redefining 
\begin{equation}
{Q}_r \to Z_r^{1/2} \left(1+\frac{1}{2} \frac{\partial \ln Z_r (X, X^\dagger, \mu)}{\partial \ln|X|} \frac{F}{M} \theta^2  \right) {Q}'_r 
\end{equation}
the soft SUSY breaking terms, defined by 
\begin{equation}
\mathcal{L}_{soft} = \frac{1}{2} (M_\lambda \lambda \lambda + \mbox{h.c.}) - \widetilde{m}^2_{Q_r} \widetilde{Q}_r^\dagger \widetilde{Q}_r - \Bigl(\sum_r A_r \widetilde{Q}_r \partial_{\widetilde{Q}_r} W(\widetilde{Q}) + \mbox{h.c.} \Bigr)
\end{equation}
can be easily extracted: 
\begin{equation}
\label{softgauginoth}
M_\lambda(\mu) = - \frac{1}{2} \frac{\partial \ln \mathcal{R} (X, \mu)}{\partial \ln |X|} \bigg\vert_{X=M} \frac{F}{M} \,,
\end{equation}
\begin{equation}
\label{softsfermionth}
\widetilde{m}^2_{Q_r}(\mu) = - \frac{1}{4} \frac{\partial^2 \ln Z_r (X, X^\dagger, \mu)}{(\partial \ln |X|)^2} \bigg\vert_{X=M} \frac{F F^\dagger}{M M^\dagger} \,,
\end{equation}
\begin{equation}
\label{softtrilinearth}
A_r(\mu) =  \frac{\partial \ln Z_r(X, X^\dagger, \mu)}{\partial \ln |X|} \bigg\vert_{X=M} \frac{F}{M} \,,
\end{equation}
where in eq.(\ref{softtrilinearth}) the permutation over the indices $(rst)$ is understood. In the following sections  we  compute explicitly the wave function  RG evolutions and extract the soft terms.

\subsection*{Coupling evolutions}
As seen in equation (\ref{softgauginoth}) the fundamental ingredient to compute gaugino masses $M$ is the knowledge of the $X$ dependence of the real coupling $\mathcal{R}$. The latter can be easily obtained once the evolution of the holomorphic coupling $S$ is known. 

Since we wish to calculate the $X$ dependence of the holomorphic coupling at low energies we have to calculate its RG evolution from the high energy scale $\Lambda_{UV}$ down to the scale $\mu$ accross the threshold $\mu_X$ of the physical messenger scale \cite{WeinbergHall},
\begin{equation}
\label{muX}
\mu_X^2 = \frac{XX^{\dagger}}{Z^2_M(\mu_X)} \,.
\end{equation}
The evolution can be split in two different regions, namely above (where quantities are denoted with the superscript $>$) and below (where quantities are denoted with the superscript $<$) the scale $\mu_X$. The coupling at $\mu_X$ is related to the high energy one by the simple formula 
\begin{equation}
S^>(\mu_X) = S^>(\Lambda) + \frac{b^>}{16 \pi^2} \, \ln \frac{\mu_X}{\Lambda} \, ,
\end{equation}
where $b = 3 T_{G} - \sum_\phi T_{\phi}$ and $\phi$ runs over all the matter fields present in the theory at a certain scale; $T_\phi$ is the Dynkin index of the representation $\phi$ ($G$ refers as usual to the adjoint representation). Writing the equivalent formula below $\mu_X$ one obtains the low energy value of the holomorphic coupling after applying matching conditions at $\mu_X$ scale, 
\begin{equation}
S(\mu) = S^>(\Lambda) + \frac{b^>}{16 \pi^2} \, \ln \frac{\mu_X}{\Lambda} + \frac{b^<}{16 \pi^2} \, \ln \frac{\mu}{\mu_X} \, .
\end{equation}
The relation between the holomorphic coupling and the interaction one is given by \cite{ArkaniHamedMurayama}
\begin{equation}
\label{real-holomorphic}
\mathcal{R}(\mu) = \re\Bigl(S(\mu)\Bigr) + \frac{T_{G}}{8 \pi^2} \, \ln \re\Bigl(S(\mu)\Bigr) - \sum_r \frac{T_{r}}{8 \pi^2} \, \ln Z_r(\mu) \,,
\end{equation}
where by $\re(S)$ we mean the real part of $S$ and the $Z_r$'s are the wave functions of the matter fields. From the previous considerations on $S$ we can easily obtain the matching condition
\begin{eqnarray}
\mathcal{R}^>(\mu_X) &=& \mathcal{R}^>(\Lambda) + \frac{b^>}{16 \pi^2} \, \ln \frac{\mu_X^2}{\Lambda^2} + \frac{T_{G}}{8 \pi^2} \, \ln \frac{\re\Bigl(S^>(\mu_X)\Bigr)}{\re\Bigl(S^>(\Lambda)\Bigr)} + \nonumber\\
&& - \sum_r \frac{T_{r}}{8 \pi^2} \, \ln \frac{Z^>_r(\mu_X)}{Z^>_r(\Lambda)} -\frac{T_{M}}{8 \pi^2} \, \ln \frac{Z_M(\mu_X)}{Z_M(\Lambda)} \, ,
\end{eqnarray}
that yields for the low energy real coupling
\begin{equation}
\mathcal{R}(\mu) = \mathcal{R}^>(\mu_X) + \frac{b_i^<}{16 \pi^2} \, \ln \frac{\mu^2}{\mu_X^2} + \frac{T_{G}}{8 \pi^2} \, \ln \frac{\re\Bigl(S^<(\mu)\Bigr)}{\re(S^<\Bigl(\mu_X)\Bigr)} - \sum_r \frac{T_{r}}{8 \pi^2} \, \ln \frac{Z^<_r(\mu)}{Z^<_r(\mu_X)} \,.
\end{equation}

\subsection*{$Z_r$ evolution}

The calculation of soft terms is just a step away: we still have to calculate the wave function renormalization dependence on $X$. Once we will have done that the whole determination of the SUSY breaking parameters will be straightforward.

The RG evolution of the matter fields is strictly connected to the knowledge of the $\gamma$ functions of the fields under consideration. In particular it is governed by the well known differential equation
\begin{equation}
\frac{d \, \ln Z_r}{dt} = \gamma_r \,.
\end{equation}
The integration of such a differential equation across the scale $\mu_X$ is the simple task to obtain the wave function $Z$ of the field $r$: indeed one gets
\begin{equation}
\label{lnZr}
\ln Z_r (\mu) = \int_{\Lambda}^{\mu_X} dt \, \gamma^>_r (t) + \int_{\mu_X}^{\mu} dt \, \gamma^<_r (t,\mu_X) \,.
\end{equation}

\subsection*{Soft terms}

All the ingredients for the determination of the soft SUSY breaking terms are now at our disposal, thus it is easy to perform the needed calculations. 

\subsubsection*{Gaugino masses}

As we saw in equation (\ref{softgauginoth}) gaugino masses are  given by
\begin{equation*}
M_\lambda (\mu) = - \frac{1}{2} \frac{\partial \, \ln \mathcal{R}(\mu)}{\partial \, \ln |X|} \frac{F}{M} \,.
\end{equation*}
It is easy to recast the derivation with respect to $\ln |X|$ in the following way:
\begin{equation*}
\frac{\partial}{\partial \, \ln |X|} =  \frac{\partial \, \ln \mu_X}{\partial \, \ln |X|}  \frac{\partial}{\partial \, \ln \mu_X} \,.
\end{equation*}
As we saw in equation (\ref{muX}) the scale $\mu_X$ is given by 
\begin{equation}
\mu_X^2 = \frac{XX^{\dagger}}{Z^2_M(\mu_X)} \,,
\end{equation}
so that
\begin{equation*}
\frac{\partial \, \ln \mu_X}{\partial \, \ln |X|} = 1 - \frac{\partial \, \ln Z_M(\mu_X)}{\partial \, \ln |X|} \sim 1 - \frac{\partial \, \ln Z_M(\mu_X)}{\partial \, \ln \mu_X} = 1 - \gamma^>_M \,.
\end{equation*}
Calculating the gaugino mass is now just a matter of bookkeeping:
{\allowdisplaybreaks \begin{align} 
M_\lambda = &- \frac{1}{2} \frac{\partial \, \ln \mu_X}{\partial \, \ln |X|}  \frac{\partial \, \ln \mathcal{R}(\mu)}{\partial \, \ln \mu_X} \frac{F}{M} = - \frac{1}{2 \mathcal{R}(\mu_X)} \Bigl(1 - \gamma^>_M\Bigr) \Biggl[\frac{b^> - b^<}{8 \pi^2} + \nonumber \\
& + \frac{T_{G}}{8 \pi^2} \frac{1}{\re\Bigl(S(\mu_X)\Bigr)}\Biggl(\frac{b^> - b^<}{8 \pi^2}\Biggr) - \sum_r \frac{T_{r}}{8 \pi^2} \Bigl(\gamma_r^> - \gamma_r^<\Bigr) - \frac{T_{M}}{8 \pi^2} \gamma^>_M\Biggr] \,.
\end{align}

\subsubsection*{Sfermion masses}
The soft masses for the low energy fields of the theory are given by equation (\ref{softsfermionth}) 
\begin{equation*}
\widetilde{m}^2_{Q_r} (\mu) = - \frac{1}{4} \frac{\partial^2 \, \ln Z_r(\mu)}{\partial \, \ln |X|^2} \frac{F F^\dagger}{M M^\dagger} \,,
\end{equation*}
thus we just have to deal with the derivatives of $Z_r$ as calculated in equation (\ref{lnZr}). The first derivative with respect to $\mu_X$ is
\begin{eqnarray}
\frac{\partial \, \ln Z_r(\mu)}{\partial \, \ln \mu_X}& = & \frac{\partial}{\partial \, \ln \mu_X} \Biggr[\int_{\Lambda}^{\mu_X} dt \, \gamma^>_r (t) + \int_{\mu_X}^{\mu} dt \, \gamma^<_r (t,\mu_X)\Biggl] \nn \\
&= & \gamma^>_r (\mu_X) - \gamma^<_r (\mu_X) + \int_{\mu_X}^{\mu} dt \, \frac{\partial \, \gamma^<_r (t,\mu_X)}{\partial \, \ln \mu_X} \,,
\end{eqnarray}
and the second one is
\begin{equation}
\frac{\partial^2 \, \ln Z_r(\mu)}{\partial \, \ln \mu_X^2} = \frac{\partial}{\partial \, \ln \mu_X} \Bigl(\gamma^>_r (\mu_X) - \gamma^<_r (\mu_X)\Bigr) -  \frac{\partial \, \gamma^<_r (t,\mu_X)}{\partial \, \ln \mu_X}\Big\vert_{\mu_X} \,.
\end{equation}
Suppose we now consider the generic coupling $\lambda$.  Its RG evolution is controlled  by the $\beta_\lambda$ function defined through
\begin{equation}
\label{evlambda}
\frac{d \, \lambda}{dt} = \beta_{\lambda} \,.
\end{equation} 
The formal  solution of eq. (\ref{evlambda}) from the high scale $\Lambda_{UV}$ down to low energies across the scale $\mu_X$ is
\begin{equation}
\lambda (\mu) = \lambda(\Lambda_{UV}) + \int_{\Lambda}^{\mu_X} dt \, \beta^>_{\lambda} (t) + \int_{\mu_X}^{\mu} dt \, \beta^<_{\lambda} (t,\mu_X) \,.
\end{equation}
By means of the above equations we can easily translate the derivation with respect to the scale $\mu_X$ to the one with respect to the running couplings of the theory using the known $\beta$ functions. In particular we obtain
\begin{equation}
\widetilde{m}^2_{Q_r} = - \frac{1}{4} \Bigl(1 - \gamma^>_M\Bigr)^2 \Biggl(\frac{\partial \Delta \gamma_r}{\partial \lambda} \beta^>_{\lambda} - \frac{\partial \gamma^<_r}{\partial \lambda} \Delta \beta_{\lambda} \Biggr) \frac{F F^\dagger}{M M^\dagger} \,,
\end{equation}
where $\Delta \gamma_r = \gamma_r^>-\gamma_r^<$ and $\Delta \beta_\lambda = \beta_\lambda^>-\beta_\lambda^<$ are defined to be the difference of the shown quantities above and below the $\mu_X$ scale.

\subsubsection*{Trilinears}
In order to complete the computation of the soft SUSY breaking terms we have to focus on the trilinears terms $A_{rst}$ of equation (\ref{lagrangianlowen}). In general the computation of the latter can be obtained through the summation over the vertex corrections owed to the different fields involved in the interaction. In particular one obtains that for any of the field of the interaction it is possible to write, as shown in (\ref{softtrilinearth}),
\begin{equation}
A_i(\mu) = \frac{\partial \, \ln Z_i(\mu)}{\partial \, \ln |X|} \frac{F}{M} \,,
\end{equation}
that following the same procedure of the previos subsection yields
\begin{equation}
A_i(\mu) = \Bigl(1 - \gamma^>_M\Bigr) \Delta \gamma_i \frac{F}{M} \,.
\end{equation}
The trilinear soft term entering the lagrangian can now be easily obtained by summing the contribution coming from any of the vertices entering the diagram, thus the resulting SUSY breaking lagrangian term will be\begin{equation}
A_{rst} \widetilde{Q}_r \widetilde{Q}_s \widetilde{Q}_t  = \Bigl(1 - \gamma^>_M\Bigr) \Biggl(\sum_{i=r,s,t} \Delta \gamma_i \, \widetilde{Q}_i \partial_{\widetilde{Q}_i} W(\widetilde{Q}) \Biggr) \frac{F}{M} \,.
\end{equation}

\subsection*{SUSY breaking contributions in our model}
\label{soft}
In the following we write all the SUSY breaking soft terms in our framework calculated using the formulae just calculated. To easily spot the numbers of loop factors at which any of the following terms arise we use $Lp = (4 \pi)^{-2}$.

\subsubsection*{Trilinears}
{\allowdisplaybreaks \begin{align} 
A_u & = \frac{Lp}{2} \, h_0^2 \, Y_u \, B_\phi \\
A_d & = \frac{Lp}{2} \, h_0^2 \, Y_d \, B_\phi \\
A_e & = \frac{Lp}{2} \, h_0^2 \, Y_e \, B_\phi 
\end{align}

\subsubsection*{$B_{\mu}$-like terms}
{\allowdisplaybreaks \begin{align} 
B_{\mu} & = Lp \, h_0^2 \, \mu \, B_\phi 
\end{align}

\subsubsection*{Gaugino masses}
{\allowdisplaybreaks \begin{align}
M_{1}  = & \, Lp^2 \, g_1^2 \, \Biggl( \frac{3}{5} h_0^2 + \frac{2}{5} h_t^2 \Biggr) \, B_\phi \\ 
M_{2}  = & \, Lp^2 \, g_2^2 \, h_0^2 \, B_\phi \\  
M_{3}  = & \, Lp^2 \, g_3^2 \, h_t^2 \, B_\phi 
\end{align}}

\subsubsection*{Soft squared masses}
{\allowdisplaybreaks \begin{align} 
{m^2_{q}}^{(2)} & = \frac{Lp^2}{2} \, h_0^2 \, \Bigl(Y_d^\dagger Y_d + Y_u^\dagger Y_u\Bigr) \, B_\phi^2 \\
{m^2_{l}}^{(2)} & = \frac{Lp^2}{2} \, h_0^2 \, Y_e^\dagger Y_e \, B_\phi^2 \\
{m^2_{u^c}}^{(2)} & = Lp^2 \, h_0^2 \, Y_u Y_u^\dagger \, B_\phi^2 \\
{m^2_{d^c}}^{(2)} & = Lp^2 \, h_0^2 \, Y_d Y_d^\dagger \, B_\phi^2 \\
{m^2_{e^c}}^{(2)} & = Lp^2 \, h_0^2 \, Y_e Y_e^\dagger \, B_\phi^2 \\
{m^2_{H_u}}^{(2)} & = \frac{Lp^2}{2} \, h_0^2 \, \Biggl[\frac{3 g_1^2}{5} + 3 g_2^2 - 4 h_0^2 - 3 h_t^2 - \tr\Bigl(Y_e^\dagger Y_e + 3 Y_d^\dagger Y_d\Bigr) - 2 \eta^2 \Biggr] \, B_\phi^2 \\
{m^2_{H_d}}^{(2)} & = \frac{Lp^2}{2} \, h_0^2 \, \Biggl[\frac{3 g_1^2}{5} + 3 g_2^2 - 4 h_0^2 - 3 h_t^2 - 3 \tr\Bigl(Y_u^\dagger Y_u\Bigr) - 2 \eta^2 \Biggr] \, B_\phi^2
\end{align}}

{\allowdisplaybreaks \begin{align} 
{m^2_{q}}^{(3)} = \, & Lp^3 \Biggl\{ h_0^2 \Biggl[ \frac{1}{50} g_1^4 + \frac{3}{2} g_2^4 - 2 \Bigl(Y_d^\dagger Y_d Y_d^\dagger Y_d + Y_u^\dagger Y_u Y_u^\dagger Y_u\Bigl) + \nn \\
& - \Bigl( \frac{1}{3}  g_1^2  + 3  g_2^2  + \frac{8}{3}  g_3^2  - 3 h_0^2 - 3 h_t^2 - 2 \eta^2 \Bigr) \Bigl(Y_u^\dagger Y_u + Y_d^\dagger Y_d\Bigr)\Biggr] +  \nn \\
& + \Bigl(\frac{1}{75}  g_1^4 + \frac{8}{3}  g_3^4\Bigr) h_t^2 \Biggr\} B_{\phi}^2 \\
{m^2_{l}}^{(3)} = \, & Lp^3 \Biggl\{ h_0^2 \Biggl[ \frac{9}{50} g_1^4  + \frac{3}{2} g_2^4 - 2 Y_e^{\dagger} Y_e Y_e^{\dagger} Y_e + \nn\\
& - \Bigl(\frac{3}{5}  g_1^2 + 3  g_2^2  - 3 h_0^2 - 3 h_t^2 - 2 \eta^2\Bigr) Y_e^{\dagger} Y_e\Biggr] + \frac{3}{25}  g_1^4 h_t^2 \Biggr\} B_{\phi}^2 \\
{m^2_{u^c}}^{(2)} = \, & Lp^3 \Biggl\{ h_0^2 \Biggl[ \frac{8}{25} g_1^4  +  2 \Bigl(Y_d Y_d^\dagger + Y_u Y_u^\dagger \Bigr) Y_u Y_u^\dagger + \nn\\
& - \Bigl( \frac{5}{3}  g_1^2 + 3 g_2^2 + \frac{16}{3} g_3^2 - 6 h_0^2 - 6 h_t^2 - 4 \eta^2 \Bigl) Y_u Y_u^\dagger \Biggr] + \nn\\
& + h_t^2 \Bigl( \frac{16}{75} g_1^4 + \frac{8}{3} g_3^4 \Bigr) \Biggr\} B_{\phi}^2 \\
{m^2_{d^c}}^{(2)} = \, & Lp^3 \Biggl\{ h_0^2 \Biggl[ \frac{2}{25}  g_1^4 - 2 \Bigl(Y_u Y_u^\dagger + Y_d Y_d^\dagger \Bigr) Y_d Y_d^\dagger + \nn\\
& - \Bigl( \frac{13}{15}  g_1^2 + 3  g_2^2 + \frac{16}{3}   g_3^2 - 6 h_0^2 - 6 h_t^2 -4 \eta^2 \Bigr) Y_d Y_d^\dagger \Biggr] + \nn\\
& + h_t^2 \Bigl(\frac{4}{75}  g_1^4 + \frac{8}{3}  g_3^4 \Bigr) \Biggr\} B_{\phi}^2 \\
{m^2_{e^c}}^{(2)} = \, & Lp^3 \Biggl\{ h_0^2 \Biggl[ \frac{18}{25} g_1^4 - 2  h_0^2  Y_e Y_e^\dagger Y_e Y_e^\dagger + \nn\\
& - \Bigl( 3 g_1^2 + 3 g_2^2 - 6 h_0^2 - 6 h_t^2 - 4 \eta^2 \Bigr) Y_e Y_e^\dagger \Biggr] + \frac{12}{25}  g_1^4 h_t^2 \Biggr\} B_{\phi}^2  \\
{m^2_{H_u}}^{(2)} = \, & Lp^3 \Biggl\{ h_0^2 \Biggr[ -\frac{201}{100}  g_1^4 - \frac{9}{10}  g_1^2 g_2^2 - \frac{9}{4}  g_2^4 - \frac{6}{5}  g_1^2 h_0^2  - 6  g_2^2 h_0^2  + 9  h_0^4 + \nn\\
& - \Bigl( \frac{7}{5}  g_1^2 + 9  g_2^2 + 6  g_3^2 - 15 h_0^2 - 9 \tr(Y_d^{\dagger} Y_d) - 3 \tr(Y_e^{\dagger} Y_e)  \Bigr) \tr(Y_d^{\dagger} Y_d) + \nn\\
& - \Bigl( \frac{9}{5}  g_1^2 + 3  g_2^2 - 5 h_0^2 + h_0^2  \tr(Y_e^{\dagger} Y_e) \Bigr) \tr(Y_e^{\dagger} Y_e) + 9 h_0^2 \tr(Y_u^{\dagger} Y_u) + \nn\\ 
& - 3 \tr(Y_e^{\dagger} Y_e Y_e^{\dagger} Y_e) + 3 \tr(Y_e^{\dagger} Y_e Y_d^{\dagger} Y_d) + 9 \tr(Y_d^{\dagger} Y_d Y_d^{\dagger} Y_d) + \nn\\
& - 9 \tr(Y_u^{\dagger} Y_u Y_u^{\dagger} Y_u) + h_t^2 \Bigl(\frac{1}{10}  g_1^2 + \frac{9}{2}  g_2^2 - 16  g_3^2 + 3  h_0^2  + 6  h_t^2 + 12 \eta^2 \Bigl) + \nn\\
& + \eta^2 \Bigl( \frac{3}{5}  g_1^2 + 3 g_2^2 + 10 h_0^2 + 8 \eta^2 \Bigr) \Biggr]  + \frac{3}{25}  g_1^4 h_t^2 \Biggr\} \\
{m^2_{H_d}}^{(2)} = \, & Lp^3 \Biggl\{ h_0^2 \Biggr[ -\frac{201}{100}  g_1^4 - \frac{9}{10}  g_1^2 g_2^2 - \frac{9}{4}  g_2^4 - \frac{6}{5}  g_1^2 h_0^2  - 6  g_2^2 h_0^2  + 9  h_0^4 + \nn\\ 
& - \Bigl( \frac{13}{5}  g_1^2 + 9  g_2^2 + 16  g_3^2 - 15  h_0^2 -  \tr(Y_u^{\dagger} Y_u) \Bigr) \tr(Y_u^{\dagger} Y_u) + \nn\\
& + 9 \tr(Y_u^{\dagger} Y_u Y_u^{\dagger} Y_u) - 9 \tr(Y_d^{\dagger} Y_d Y_d^{\dagger} Y_d) - 3 \tr(Y_e^{\dagger} Y_e Y_e^{\dagger} Y_e) +\nn \\
& + 9 h_0^2  \tr(Y_d^{\dagger} Y_d) + 3 h_0^2  \tr(Y_e^{\dagger} Y_e) + h_t^2 \Bigl( \frac{1}{10}  g_1^2 + \frac{9}{2}  g_2^2 - 16  g_3^2 + 3 h_0^2 + \nn\\
& + 6 h_t^2 \Bigr) + \frac{3}{5}   g_1^2   \eta^2 + 3  g_2^2  \eta^2 + 10  h_0^2  \eta^2 + 12 h_t^2  \eta^2 + 8 \eta^4 \Biggr] + \frac{3}{25}  g_1^4 ht^2 \Biggr\} 
 \end{align}}


\begin{thebibliography}{30}

\bibitem{susy}
S.~Dimopoulos and H.~Georgi,
  ``Softly Broken Supersymmetry And SU(5),''
  Nucl.\ Phys.\ B {\bf 193}, 150 (1981);
  S.~Dimopoulos, S.~Raby and F.~Wilczek,
  ``Supersymmetry And The Scale Of Unification,''
  Phys.\ Rev.\ D {\bf 24}, 1681 (1981).


\bibitem{ATLAS}
ATLAS collaboration, \\
https://twiki.cern.ch/twiki/bin/view/AtlasPublic/SupersymmetryPublicResults.

\bibitem{CMS}
CMS collaboration, \\
https://twiki.cern.ch/twiki/bin/view/CMSPublic/PhysicsResults.

\bibitem{PDG}
K.~Nakamura {\it et al.} [ Particle Data Group Collaboration ],
J.\ Phys.\ G {\bf G37}, 075021 (2010).

  \bibitem{DMall}
Xenon collaboration,  http://xenon.physik.uni-mainz.de/papers.html;
FERMI-LAT collaboration, http://www-glast.stanford.edu/;
  V.~Bertin [ ANTARES Collaboration ],
  J.\ Phys.\ Conf.\ Ser.\  {\bf 315}, 012030 (2011);
ICECUBE collaboration, http://icecube.wisc.edu/;
  G.~Bertone, (ed.),
  Cambridge, UK: Univ. Pr. (2010) 738 p;
  B.~Censier,
   [arXiv:1110.0191 [astro-ph.CO]].
  

\bibitem{2loopgeneral}
  M.~-x.~Luo, H.~-w.~Wang, Y.~Xiao,
  Phys.\ Rev.\  {\bf D67}, 065019 (2003).
  [hep-ph/0211440].

\bibitem{2loopSM}
  M.~-x.~Luo, Y.~Xiao,
  Phys.\ Rev.\ Lett.\  {\bf 90}, 011601 (2003).
  [hep-ph/0207271].
  

\bibitem{ArkaniHamed:2004fb}
  N.~Arkani-Hamed, S.~Dimopoulos,
  JHEP {\bf 0506}, 073 (2005).
  [hep-th/0405159];
  N.~Arkani-Hamed, S.~Dimopoulos, G.~F.~Giudice, A.~Romanino,
  Nucl.\ Phys.\  {\bf B709}, 3-46 (2005).
  [arXiv:hep-ph/0409232 [hep-ph]].
  
  
 
\bibitem{Giudice:2011cg}
  G.~F.~Giudice and A.~Strumia,
  arXiv:1108.6077 [hep-ph].
  
  
\bibitem{Alves:2011ug}
  D.~S.~M.~Alves, E.~Izaguirre, J.~G.~Wacker,
    [arXiv:1108.3390 [hep-ph]];
  J.~Cao, W.~Wang, J.~M.~Yang,
   [arXiv:1108.2834 [hep-ph]].

  
\bibitem{Chacko:2001km}
  Z.~Chacko, E.~Ponton,
  Phys.\ Rev.\  {\bf D66}, 095004 (2002).
  [hep-ph/0112190];
  Z.~Chacko, E.~Katz, E.~Perazzi,
  Phys.\ Rev.\  {\bf D66}, 095012 (2002).
  [hep-ph/0203080];
  F.~R.~Joaquim, A.~Rossi,
  Nucl.\ Phys.\  {\bf B765}, 71-117 (2007).
  [hep-ph/0607298].
  
  
\bibitem{Giudice:1998bp}
G.~F.~Giudice, R.~Rattazzi,
Phys.\ Rept.\  {\bf 322 } (1999)  419-499.
[hep-ph/9801271].

  
\bibitem{Endo:2010fk}
  M.~Endo, S.~Shirai, T.~T.~Yanagida,
  Prog.\ Theor.\ Phys.\  {\bf 125 } (2011)  921-932.
  [arXiv:1009.3366 [hep-ph]].

\bibitem{inprogress}
  F.~Bazzocchi, M.~Monaco,
 in progress.


\bibitem{anomaly}
  G.~F.~Giudice, M.~A.~Luty, H.~Murayama, R.~Rattazzi,
  JHEP {\bf 9812}, 027 (1998).
  [hep-ph/9810442];
 

\bibitem{RGE}
  S.~P.~Martin,
  In *Kane, G.L. (ed.): Perspectives on supersymmetry* 1-98.
  [arXiv:hep-ph/9709356 [hep-ph]];
  S.~P.~Martin, M.~T.~Vaughn,
  Phys.\ Rev.\  {\bf D50}, 2282 (1994).
  [hep-ph/9311340].
  

\bibitem{Okada:1990gg}
  Y.~Okada, M.~Yamaguchi, T.~Yanagida,
  Phys.\ Lett.\  {\bf B262}, 54-58 (1991).

\bibitem{ColemanWeinberg}
  S.~R.~Coleman, E.~J.~Weinberg,
  Phys.\ Rev.\  {\bf D7}, 1888-1910 (1973).


\bibitem{Casas_Quiros}
  J.~A.~Casas, J.~R.~Espinosa, M.~Quiros, A.~Riotto,
  Nucl.\ Phys.\  {\bf B436}, 3-29 (1995).
  [arXiv:hep-ph/9407389 [hep-ph]].
  

\bibitem{TSU1}
B.~Holdom and J.~Terning, Phys. Lett. {\bf B247} {(1990)} 88--92;
M.~E. Peskin and T.~Takeuchi, Phys. Rev. Lett. {\bf 65} {(1990)} 964--967;
M.~Golden and L.~Randall, Nucl. Phys. {\bf B361} {(1991)} 3--23;
A.~Dobado, D.~Espriu, and M.~J. Herrero, Phys. Lett. {\bf B255} {(1991)}
  405--414;
M.~E. Peskin and T.~Takeuchi, Phys. Rev. {\bf D46} {(1992)} 381--409;

\bibitem{TSU2}
I.~Maksymyk, C.~P. Burgess, and D.~London, Phys. Rev. {\bf D50} {(1994)}
  529--535 {[arXiv: hep-ph/9306267]}.


\bibitem{Drees:1990dx}
  M.~Drees, K.~Hagiwara,
  Phys.\ Rev.\  {\bf D42}, 1709-1725 (1990).


\bibitem{Giudice:2008uk}
  G.~F.~Giudice, M.~Nardecchia, A.~Romanino,
  Nucl.\ Phys.\  {\bf B813}, 156-173 (2009).  
  
\bibitem{FroggattNielsen}
  C.~D.~Froggatt, H.~B.~Nielsen,
  Nucl.\ Phys.\  {\bf B147}, 277 (1979).  
  
\bibitem{Buras:1997ij}
  A.~J.~Buras, A.~Romanino, L.~Silvestrini,
  Nucl.\ Phys.\  {\bf B520}, 3-30 (1998).
  [hep-ph/9712398].
  [arXiv:0812.3610 [hep-ph]].
  
\bibitem{Atwood:1996vj}
  D.~Atwood, L.~Reina, A.~Soni,
  Phys.\ Rev.\  {\bf D55}, 3156-3176 (1997).
  [hep-ph/9609279].
  
\bibitem{Buras:2010mh}
  A.~J.~Buras, M.~V.~Carlucci, S.~Gori, G.~Isidori,
  JHEP {\bf 1010}, 009 (2010).
  [arXiv:1005.5310 [hep-ph]].  


\bibitem{BBN}
  D.~Tytler, J.~M.~O'Meara, N.~Suzuki and D.~Lubin,
  Phys.\ Scripta {\bf T85}, 12 (2000)
  [arXiv:astro-ph/0001318].

  
\bibitem{feng}
  J.~L.~Feng,
  Ann.\ Rev.\ Astron.\ Astrophys.\  {\bf 48}, 495 (2010).
  [arXiv:1003.0904 [astro-ph.CO]];
 

\bibitem{FileviezPerez:2009im}
  P.~Fileviez Perez, H.~Iminniyaz, G.~Rodrigo, S.~Spinner,
  Phys.\ Rev.\  {\bf D81}, 095013 (2010).
  [arXiv:0911.1360 [hep-ph]].


\bibitem{Hirsch:2000ef}
  M.~Hirsch, M.~A.~Diaz, W.~Porod, J.~C.~Romao, J.~W.~F.~Valle,
  Phys.\ Rev.\  {\bf D62}, 113008 (2000).
  [hep-ph/0004115].

\bibitem{Ross:1984yg}
  G.~G.~Ross, J.~W.~F.~Valle,
  Phys.\ Lett.\  {\bf B151}, 375 (1985);
  A.~Santamaria, J.~W.~F.~Valle,
  Phys.\ Lett.\  {\bf B195}, 423 (1987); 
  A.~Masiero, J.~W.~F.~Valle,
  Phys.\ Lett.\  {\bf B251}, 273-278 (1990); 
  J.~C.~Romao, J.~W.~F.~Valle,
  Nucl.\ Phys.\  {\bf B381}, 87-108 (1992); 
  M.~Hirsch, W.~Porod, M.~A.~Diaz, J.~C.~Romao, J.~W.~F.~Valle,
   [hep-ph/0202149];
  M.~A.~Diaz, M.~Hirsch, W.~Porod, J.~C.~Romao, J.~W.~F.~Valle,
  Phys.\ Rev.\  {\bf D68}, 013009 (2003).
  [hep-ph/0302021];
  A.~Vicente Montesinos, and references therein.
  [arXiv:1104.0831 [hep-ph]].
  

\bibitem{Restrepo:2011rj}
  D.~Restrepo, M.~Taoso, J.~W.~F.~Valle, O.~Zapata,
    [arXiv:1109.0512 [hep-ph]].


\bibitem{2loop}
S.~P.~Martin and M.~T.~Vaughn,
Phys.\ Rev.\  D {\bf 50}, 2282 (1994)
[Erratum-ibid.\  D {\bf 78}, 039903 (2008)]
[arXiv:hep-ph/9311340].
 
  
\bibitem{Giudice:1997ni}
  G.~F.~Giudice, R.~Rattazzi,
  Nucl.\ Phys.\  {\bf B511}, 25-44 (1998).
  [hep-ph/9706540];
  N.~Arkani-Hamed, G.~F.~Giudice, M.~A.~Luty, R.~Rattazzi,
  Phys.\ Rev.\  {\bf D58}, 115005 (1998).
  [hep-ph/9803290].

\bibitem{ArkaniHamedMurayama}
 ÊN.~Arkani-Hamed, H.~Murayama,
 Ê
  JHEP {\bf 0006 } (2000) Ê030,
 Ê[hep-th/9707133];
 ÊN.~Arkani-Hamed, H.~Murayama,
 Ê
 ÊPhys.\ Rev.\ Ê{\bf D57 } (1998) Ê6638-6648,
 Ê[hep-th/9705189].

\end{thebibliography}
\end{document}